\documentclass[referee]{jfm}
\usepackage{graphicx}
\usepackage{natbib}

\let\realverbatim=\verbatim
\let\realendverbatim=\endverbatim
\renewcommand\verbatim{\par\addvspace{6pt plus 2pt minus 1pt}\realverbatim}
\renewcommand\endverbatim{\realendverbatim\addvspace{6pt plus 2pt minus 1pt}}
\makeatletter
\newcommand\verbsize{\@setfontsize\verbsize{10}\@xiipt}
\renewcommand\verbatim@font{\verbsize\normalfont\ttfamily}
\makeatother

\ifCUPmtlplainloaded \else
  \checkfont{eurm10}
  \iffontfound
    \IfFileExists{upmath.sty}
      {\typeout{^^JFound AMS Euler Roman fonts on the system,
                   using the 'upmath' package.^^J}%
       \usepackage{upmath}}
      {\typeout{^^JFound AMS Euler Roman fonts on the system, but you
                   don't seem to have the}%
       \typeout{'upmath' package installed. jfm.cls can take advantage
                 of these fonts,^^Jif you use 'upmath' package.^^J}%
      }
  \else
  \fi
\fi

\ifCUPmtlplainloaded \else
  \checkfont{msam10}
  \iffontfound
    \IfFileExists{amssymb.sty}
      {\typeout{^^JFound AMS Symbol fonts on the system, using the
                'amssymb' package.^^J}%
       \usepackage{amssymb}%

      }{}
  \fi
\fi

\ifCUPmtlplainloaded \else
  \IfFileExists{amsbsy.sty}
    {\typeout{^^JFound the 'amsbsy' package on the system, using it.^^J}%
     \usepackage{amsbsy}}
    {}
\fi

\newsavebox{\astrutbox}
\sbox{\astrutbox}{\rule[-5pt]{0pt}{20pt}}

\title[Plugs in rough capillary tubes]{Plugs in rough capillary tubes: enhanced dependence of motion
on plug length}

\author[Q. Zhang, K. S. Turitsyn, and T. A. Witten]{\ns Q. \ns Z\ls H\ls A\ls N\ls G$^1$,\ns
K. \ns S. \ns T\ls U\ls R\ls I\ls T\ls S\ls Y\ls N$^{1,2,3}$,\ns
\and T.\ns A.\ns W\ls I\ls T\ls T\ls E\ls N$^1$}

\affiliation{$^1$Department of Physics and the James Frank
Institute, University of Chicago, Chicago, IL 60637, USA \\[\affilskip]
$^2$ T-4 \& CNLS, Los Alamos National Laboratory, Los Alamos, NM
87545, USA \\[\affilskip]
$^3$ Landau Institute for Theoretical Physics, Moscow, 119334,
Russia}

\date{10 March 2010}
\pubyear{} 
\volume{}
\pagerange{\pageref{firstpage}--\pageref{lastpage}}
\doi{S002211200100456X}

\begin{document}

\label{firstpage}
\maketitle

\begin{abstract}
We discuss the creeping motion of plugs of negligible viscosity in
rough capillary tubes filled with carrier fluids. This extends
Bretherton's research work on the infinite-length bubble motion in a
cylindrical or smooth tube for small capillary numbers $Ca$
\cite{Bretherton_1961}. We first derive the asymptotic dependence of
the plug speed on the finite length in the smooth tube case. This
dependence on length is exponentially small, with a decay length
much shorter than the tube radius $R$. Then we discuss the effect of
azimuthal roughness of the tube on the plug speed. The tube
roughness leads to an unbalanced capillary pressure and a carrier
fluid flux in the azimuthal plane. This flux controls the relaxation
of the plug shape to its infinite-length limit. For long-wavelength
roughness, we find that the above decay length is much longer in the
rough tube, and even becomes comparable to the tube radius $R$ in
some cases. This implies a much-enhanced dependence of the plug
speed on the plug length. This mechanism may explain the catch-up
effect seen experimentally \cite{Ying_2008}.
\end{abstract}


\section{Introduction}
\indent The field of multiphase microfluidics has expanded greatly
in the last decade because of the numerous applications of the
technique in chemistry and biology \cite{Gunther_2006}. Transporting
reactants through microchannels in the form of bubbles or droplets
has many advantages over the traditional single-phase systems. These
include enhanced mixing rate, reduced dispersion and higher
interface areas \cite{Gunther_2006}. However there are still many
technological problems that remain unsolved. A particular problem
that aroused our attention is the control of spacing within a train
of long droplets or plugs carried along the channel by a wetting
fluid. Such plugs are separated from the tube by a thin lubricating
film, whose thickness is controlled by the speed of motion
\cite{Bretherton_1961}. It is commonly observed \cite{Ying_2008}
that the separation between the plugs changes as they move.
Eventually one plug can coalesce with its neighbor, a process known
as the catch-up effect. To our knowledge there is no widely accepted
explanation of this effect. \\

\indent A change in the distance between neighboring plugs is
possible only if the fluid flux in the lubricating layer is
different in the two plugs. Thus in order to understand the reason
for the catch-up effect one needs to analyze the shape and dynamics
of this lubricating layer. There are many physical mechanisms that
could be responsible for the structure of the lubricating layer. In
this work we analyze only one of them: irregularities of the channel
shape. We argue that these irregularities, \textit{i.e.} roughness
of the tube wall, could be responsible for the fluctuations of the
distance between the plugs. The influence of the substrate geometry
or roughness on the lubricating layer thickness has been recognized
for several decades and has been studied quite extensively
(\cite{Stillwagon_1988}; \cite{Schwartz_1995};
\cite{Kalliadasis_2000}; \cite{Mazouchi_2001}; \cite{Howell_2003}).
In the present work we consider an effect beyond the direct affect
on speeds. We find that the roughness enhances dramatically the
sensitivity of the thin film width to the plug length, and thus
produces speed variations between individual plugs that might
eventually lead to catch-up coalescence events.\\

\indent The earliest discussion of the roughness effect on the plug
motion inside capillary tubes dates back to the seminal paper of
Bretherton \cite{Bretherton_1961}. He suggested that the roughness
can be important if its amplitude is comparable to or larger than
the lubricating layer thickness. Therefore the roughness effect is
usually negligible for large enough tubes where other physical
mechanisms are more important (see \textit{e.g.}
\cite{Gunther_2006}, \cite{Ajaev_2006} for review). However as
modern microfluidic channels become smaller, surface roughness
becomes more important relative to the channel width. Probably the
first experimental observation that is attributed to the effect of
roughness is reported in the work of Chen \cite{Chen_1986}, where
deviations from the Bretherton's predictions were observed for small
capillary numbers $Ca$. \\

\indent On the theoretical side the most widely studied effect of
the surface roughness corresponds to the limit where the roughness
amplitude is significantly smaller than the lubricating film
thickness. In this case the roughness effect can be accounted for
via effective boundary conditions: the rough surface can be modeled
as a smooth one with non-zero slip length determined by the
roughness properties (\cite{Einzel_1990,Miksis_1994,DeGennes_2002}).
Krechetnikov and Homsy \cite{Krechetnikov_2005} have analyzed the
effect of the slip length on the lubricating film profile, and
showed that the effect is negligible for the slip length smaller
than the Bretherton thin film thickness, but leads to a dramatic
increase of the lubricating film thickness in the opposite limit: in
this case the lubricating film thickness becomes comparable to the
slip length and is independent of $Ca$. In a broader context there
have been a numerous studies on how the channel geometry affects the
dynamics and shape of plugs. Wong et al. (\cite{Wong_1995};
\cite{Wong_1995_2}) considered the plug motion in polygonal channels
and showed that the shape of the plug strongly depends on the
capillary number and that in the limit $Ca \ll 1$, the fluid flux in
the lubricating film does not depend on $Ca$. Hazel and Heil
\cite{Heil_2002} confirmed this effect by numerical simulations, and
showed that a similar behavior is observed in tubes with elliptical
cross-sections as well. It should also be noted that the surface
roughness can have a non-hydrodynamic effect on the plug motion by
changing wetting properties of the surface (\cite{Wenzel_1936,
Herminghaus_2000, Bico_2001}) or dynamical destabilization of the
lubricating layer due to fluctuations.\\

\indent In this work we consider capillary tubes with roughness
smaller than the lubricating film thickness. In this regime the
roughness cannot produce any significant effect on the width of the
lubricating layer. However, as we will show, in the rough tube the
decay length of the plug shape to its infinite-length limit is
increased dramatically. In some cases, the decay length becomes
comparable to or even larger than the tube radius. The ripples of
the plug shape propagate from the front region to the rear region of
the plug, leading to a much-enhanced dependence of the fluid flux on
the plug length. On the level of the plug train this effect
translates the variations of plug speeds to the fluctuations of the
distance between them. Thus it can be responsible for the catch-up
events. We also note that this is essentially a non-local effect. To
our knowledge this effect has not been discussed perviously. \\

\indent We will illustrate this effect by using a very simple
system, a single air plug of finite length moving through an almost
cylindrical tube. For the sake of simplicity, throughout the paper
we neglect the effects of gravity, inertia and surfactant
concentrations, assuming that the only relevant stresses are due to
surface tension and viscosity. We also limit our analysis to the
small capillary number regime: $Ca \ll 1$. We assume that the tube
roughness can be modeled as weakly non-circular cross sections of
the tube, which does not change in the longitudinal direction and
can be described by a roughness function $e(\theta)$ with $\theta$
being the azimuthal angle. We will consider the situation where the
deformation is smooth enough for the lubrication approximation to be
valid, that is, $e(\theta) \ll R \cdot Ca^{2/3}$, \textit{i.e.} the
deformation is small compared to the Bretherton's thin film
thickness. However, we expect that our scaling results can be safely
extrapolated to the regime: $e(\theta) \sim R\cdot Ca^{2/3}$. \\

\indent The structure of this paper is as follows. We start our
analysis by revisiting the classical Bretherton solution for the
semi-infinite air plug moving in the perfectly cylindrical tube.
Then we follow the finite-plug analysis of Teletzke
\cite{Teletzke_1983} and extend it to a wider range of plug length.
The main finding of this section is the exponential suppression of
the speed corrections due to the plug length variations: $\delta U/U
\propto \exp\left[- L_p/(2L_\infty)\right]$, where $L_p$ measures
the length of the plug, and $L_\infty =0.643 R \cdot (3Ca)^{1/3}$ is
the characteristic decay length of the plug shape to its
infinite-length limit for the smooth tube case. In Section 3 we
explain that the presence of small tube roughness produces a whole
spectrum of relaxation modes with different decay lengths
$\lambda_i$. The relaxation mode with the decay length
$\lambda_{max}\simeq L_p$ provides the largest contribution to the
dependence on the plug length. In Section 4 we estimate of the
amplitude of the plug shape distortion due to the roughness. Then we
estimate the effect of tube roughness on the finite plug speed in
Section 5. We conclude in Section 6 by discussing the limitations of
the model and proposing future research directions. \\

\section{Plug speed dependence on the plug length in smooth tubes}

\indent Suppose we have a smooth cylindrical capillary tube with
radius $R$. It is initially filled with a liquid of viscosity $\mu$,
which is called the carrier fluid in the following discussion. Now
an air plug with negligible viscosity is forced into the tube and
moves slowly together with the carrier fluid due to a pressure
gradient along the tube. As discussed in detail later, the plug
speed $U$ is different from the speed $V$ of the carrier fluid far
away from the plug \cite{Jensen_2004}. In this section, we will
investigate the effect of the plug length on the plug speed $U$.
Both gravitational and inertial effects are assumed to be
negligible. \\
\begin{figure}
\begin{center}
\includegraphics[angle=-90, width=5.0in]{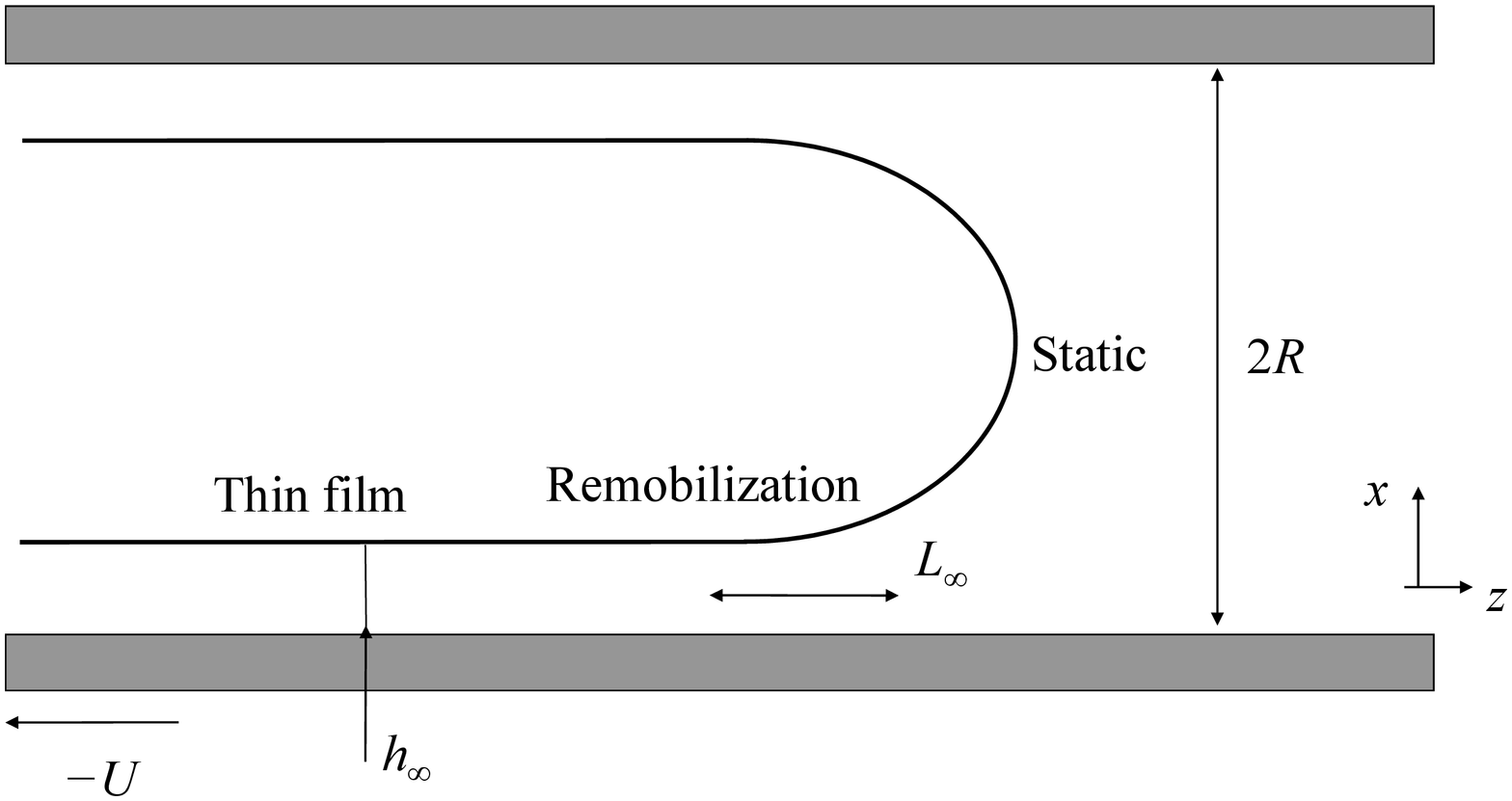}
\caption{Sketch of a plug identifying Bretherton's quantities used
in the text.} \label{fig:1:Brethertonshape}
\end{center}
\end{figure}
\subsection{Bretherton-Teletzke mechanism}

\indent \indent In Bretherton's seminal work \cite{Bretherton_1961},
he studied the limit case of semi-infinite air plug motion. He
showed that a dimensionless number $Ca$, called capillary number,
controls the plug motion:
\begin{equation}
Ca = \frac{\mu U} {\gamma} \, ,
\end{equation}
\noindent where $\gamma$ is an interfacial tension between the plug
and the carrier fluid. Bretherton's steady state solution of a
semi-infinite plug is illustrated in
Fig.\ref{fig:1:Brethertonshape}. For convenience, we have chosen the
plug as the frame of reference. In this frame, the tube wall moves
with velocity $-U$. The shape of a semi-infinite plug consists of
three regions: a spherical cap (static region) is connected to a
thin film region of almost constant liquid film thickness $h_\infty$
via a transition remobilization region. Since the internal viscosity
of the plug is zero, there can be no shear stress at the plug
surface. Thus the carrier fluid in the thin film region moves
without shear at the wall speed $U$. Bretherton also found that the
lubricating film thickness decays exponentially with a decay length
$h_\infty \, (3Ca)^{-1/3}$, denoted as $L_\infty$
\cite{Bretherton_1961}. Thus $L_\infty$ is a measure of the length
of the remobilization region. Both capillary force and viscous force
are important in the remobilization region, where the plug shape is
determined by an equilibrium between these two forces. In the limit
of small capillary numbers, $Ca < 0.005 \ll 1$, Bretherton derived
the dependence of $h_\infty$ on $Ca$ and $R$ \cite{Bretherton_1961}:
\begin{equation}
h_\infty = 0.643 \, R \cdot (3Ca)^{2/3} \, . \label{eqn:1:Linf&hinf}
\end{equation}
\noindent Obviously, both $L_\infty$ and $h_\infty$ are much less
than $R$ for small capillary numbers. Relative motion between the
plug and the carrier fluid is governed by an incompressibility
constraint. This constraint states that, in the steady state, the
total volume of the plug and the carrier fluid between two arbitrary
cross sections along the tube doesn't change over time.
Mathematically, it can be translated into the equation: $\pi R^2 \,
V = \pi (R-h_\infty)^2 \, U$. By keeping only the lowest order
approximation, we get:
\begin{eqnarray}
U / V &\simeq& 1 + 2 h_\infty /R \nonumber \\
&=& 1 + 1.286 \, (3Ca)^{2/3} \, . \label{eqn:1:BreUV}
\end{eqnarray}
\noindent Strictly speaking, Eqn.(\ref{eqn:1:BreUV}) is only valid
in the limit of the semi-infinite plug. For a finite plug, we expect
Eqn.(\ref{eqn:1:BreUV}) to be modified as \cite{Teletzke_1983}: $U/V
= 1 + 2h^*/R$, with $h^*$ different from its limiting value
$h_\infty$. Physically speaking, $-h^*$ represents the average
carrier fluid flux per unit circumference. As shown in detail later,
$h^*$ depends on the plug length and approaches $h_\infty$ as we
increase the length. \\
\begin{figure}
\begin{center}
\includegraphics[angle=-90, width=5.0in]{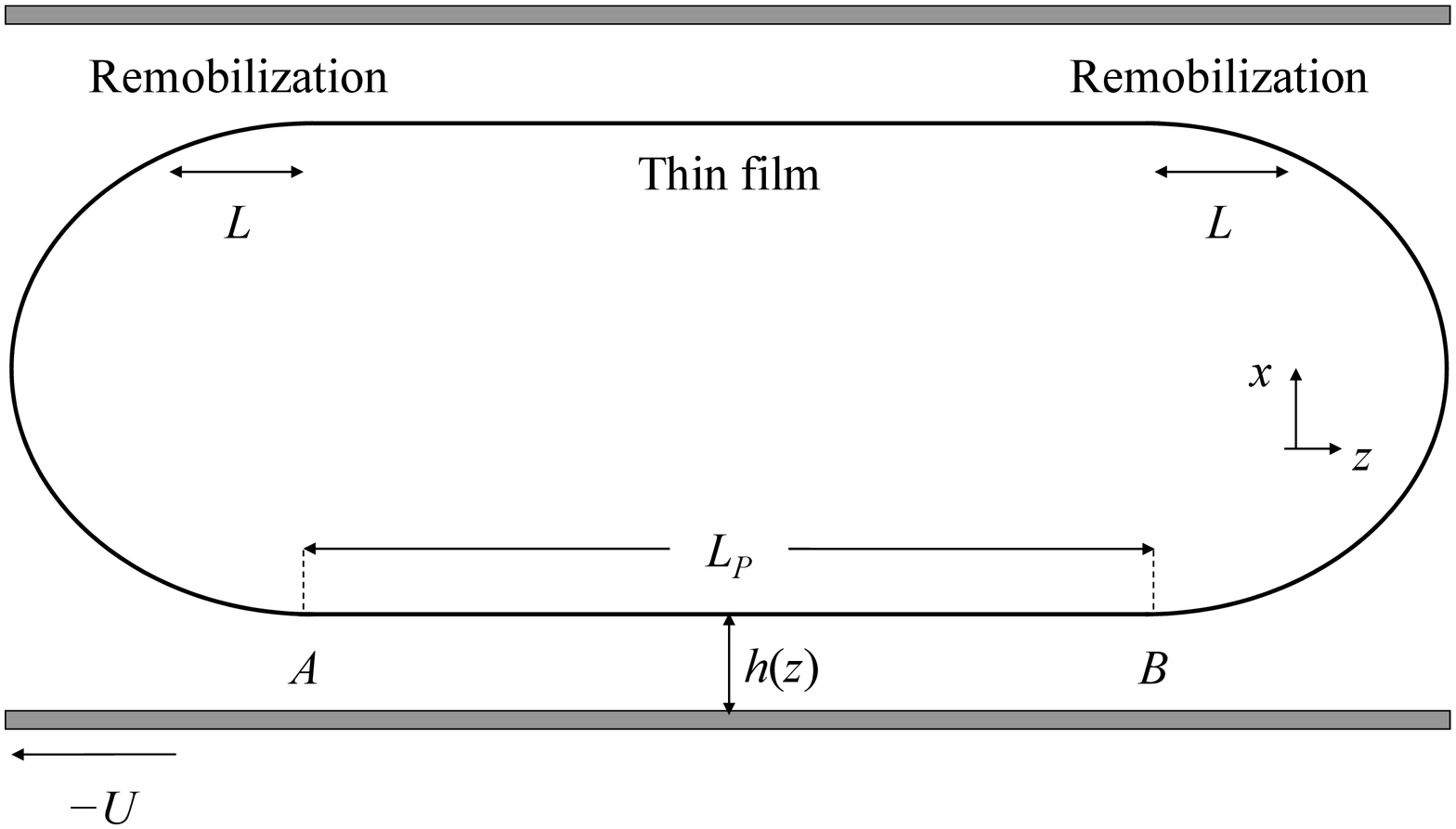}
\caption{Quantities characterizing a plug of finite length.}
\label{fig:1:finiteshape}
\end{center}
\end{figure}

\begin{figure}
\begin{center}
\includegraphics[angle=0, width=5.0in]{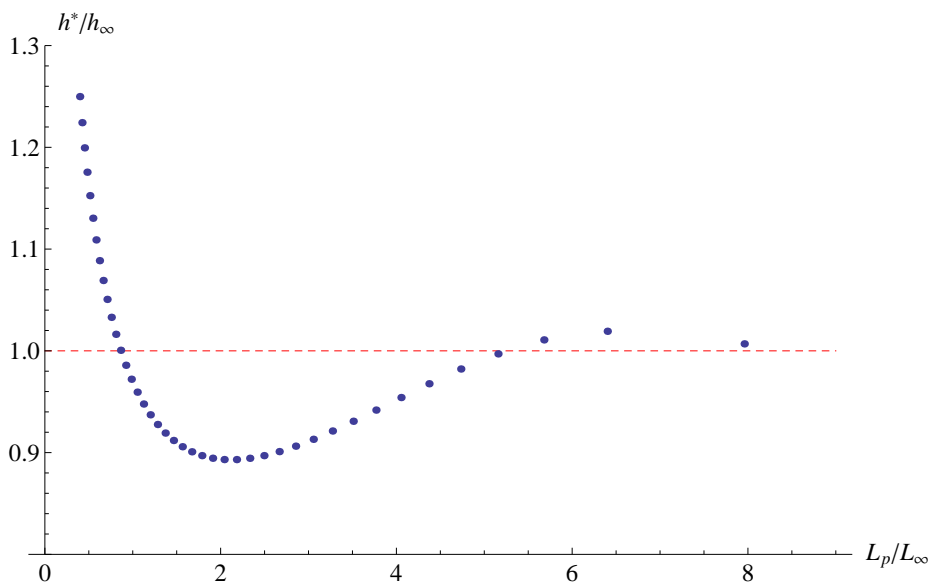}
\caption{Dependence of $h^{*}$ on the plug length $L_p$ obtained by
using Teletzke's method.} \label{fig:1:Teletzke}
\end{center}
\end{figure}

\indent The motion of a finite-length air plug in a capillary tube
was first studied by Teletzke \cite{Teletzke_1983}. In the following
discussion length, velocity, and pressure are made dimensionless by
using tube radius $R$, plug speed $U$, and capillary pressure
$\gamma/R$. Capillary number was kept small in Teletzke's analysis:
$Ca \ll 1$. By imposing a lubrication approximation and appropriate
boundary conditions (non-slip condition at the tube wall and
stress-free condition at the plug surface), the balance of viscous
and capillary pressure gradients yields \cite{Teletzke_1983}:
\begin{equation}
\frac{d^3 h}{dz^3} = \frac{3Ca \cdot (h - h^{*})}{h^3} \, ,
\label{eqn:1:shapeODE}
\end{equation}
\noindent wherever $|dh/dz| \ll 1$. For small $Ca$, this includes
both the thin film region and the remobilization region. For a plug
of fixed length, $h^*$ is a self-consistent parameter whose value is
determined by the fact that remobilization regions must match with
spherical caps at the plug ends. Suppose the front and rear
remobilization regions are separated by a distance $L_p$ for a
finite plug. Teletzke proposed a method to solve the above
differential equation numerically in the limit of short plugs: $L_p
\lesssim 10L_\infty$ \cite{Teletzke_1983}. Numerical results
obtained by using Teletzke's method are plotted in
Fig.\ref{fig:1:Teletzke}. As expected, $h^*$ approaches $h_\infty$
as the plug length increases. However, the trend is not monotonic.
The origin of this oscillating behavior will be clear in the next
section. \\

\indent Teletzke's method did not address the asymptotic behavior
for finite plugs with length: $L_p \gg L_\infty$. So the knowledge
of the dependence of the plug speed on the plug length is incomplete
at the current stage. We only understand two limit cases:
Bretherton's semi-infinite plugs and Teletzke's short plugs with
$L_p \lesssim 10L_\infty$. In the next section, we complete this
previously unstudied part of the plug speed problem, by determining
the asymptotic behavior of $h^*$ for plugs with large $L_p$. \\

\subsection{Asymptotic corrections to the semi-infinite plug speed}

\indent \indent Due to rotational symmetry, the problem of
determining plug shape is reduced to calculating a lubricating film
thickness . The governing equation of $h(z)$ is given by
Eqn.(\ref{eqn:1:shapeODE}). For small capillary numbers $Ca$, this
differential equation is valid in both the thin film and the
remobilization region, as noted above. Its solution has to match
spherical caps at the ends of remobilization regions. By using the
rescalings: $h = h^{*} \cdot \eta$ and $z = L \cdot \xi$, we may
rewrite Eqn.(\ref{eqn:1:shapeODE}) as:
\begin{equation}
\frac{d^3 \eta}{d\xi^3} = \frac{\eta - 1}{\eta^3} \, .
\label{eqn:1:newshapeODE}
\end{equation}
\noindent In the above substitutions, $L$ is the remobilization
length of the finite plug defined as: $L \equiv
h^{*}\,(3Ca)^{-1/3}$. Comparing with Eqn.(\ref{eqn:1:Linf&hinf}), we
observe that $\eta$ measures thickness in units comparable to the
asymptotic thin film thickness $h_\infty$. Likewise $\xi$ evidently
measures lengths in units comparable to the remobilization length
$L_\infty$. We denote the plug length $L_p$ in these units as
$\xi_p$. So in this section
we will study long plugs with $\xi_p \gg 1$.\\

\indent As we approach the thin film region, $\eta$ becomes
arbitrarily close to 1: $|\eta - 1| \ll 1$. So we can simplify
Eqn.(\ref{eqn:1:newshapeODE}) as follows: $d^3 \eta / d\xi^3 \simeq
\eta - 1$. This autonomous equation has solutions of the form: $\eta
- 1 = \exp{(\nu \xi)}$, where $\nu^3 = 1$. It has three roots $\nu_1
= 1$ and $\nu_{2,3} = -1/2\pm \sqrt{3}i/2$, which correspond to
three solution modes given below:
\begin{eqnarray}
\psi_1(\xi) &=& \exp{(\xi)} \, ,\nonumber \\
\psi_2(\xi) &=& \exp{(-\xi/2)}\, \cos{(\sqrt{3}\xi /2)} \, ,\nonumber \\
\psi_3(\xi) &=& \exp{(-\xi/2)}\, \sin{(\sqrt{3}\xi /2)} \, .
\end{eqnarray}
\noindent So general solutions are linear combinations of these
modes, and we have:
\begin{equation}
\eta = 1 + \sum_{i=1}^3 C_i \psi_i(\xi) \, ,
\label{eqn:1:generalsolution}
\end{equation}
\noindent where coefficients $C_i$ are determined from boundary
conditions away from the thin film region. We note for future
reference that for large positive $\xi$ , $\eta$ is dominated by the
growing $\psi_1$ function. \\

\indent Far from the thin film region treated above, the following
relation is valid: $\eta = h/h^* \gg 1$. For very small $Ca$,
lubrication approximation ($|dh/dz \ll 1|$) is still applicable in
this regime \cite{Teletzke_1983}. So Eqn.(\ref{eqn:1:newshapeODE})
can be approximated as follows: $d^3 \eta / d\xi^3 \simeq 0  $.
Integration of this equation yields parabolic profiles:
\begin{equation}
\eta^{F,R} = \frac{1}{2}P^{F,R} \, \xi^2 + Q^{F,R} \, \xi + S^{F,R}
\, , \label{eqn:1:parabolas}
\end{equation}
\noindent where superscripts $F$ and $R$ stand for front and rear
solutions, respectively. As defined by Teletzke
\cite{Teletzke_1983}, axial ($z$-direction) positions of $A$ and $B$
in Fig.\ref{fig:1:finiteshape}, correspond to axial positions of the
lowest points of front and rear parabolas. And he defined the
distance between $A$ and $B$ as the plug length $L_p$. We will keep
Teletzke's definition in the following discussion. \\

\begin{figure}
\begin{center}
\includegraphics[angle=0, width=5.0in]{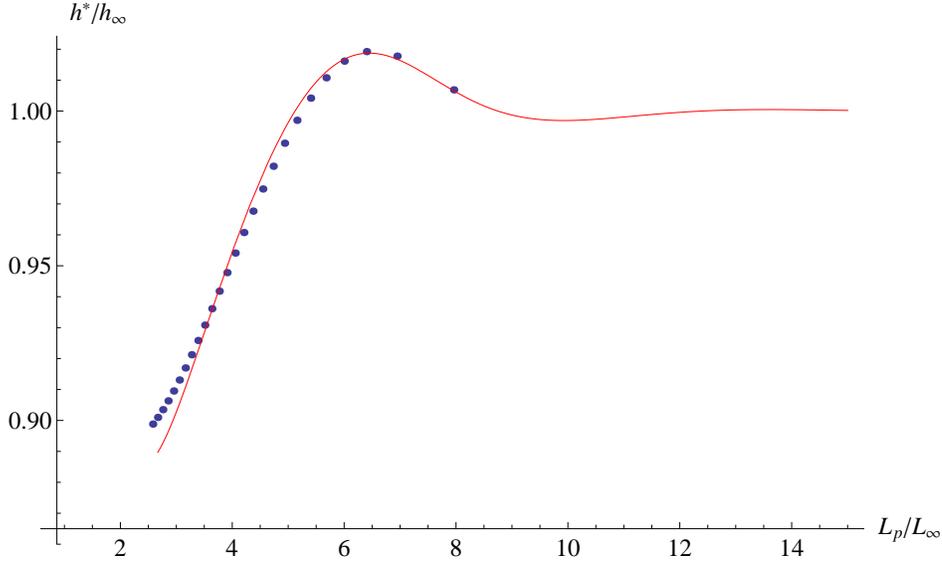}
\caption{Dependence of $h^{*}$ on the plug length $L_p$. Solid line
represents our data obtained via a perturbation analysis. Points are
data obtained via Teletzke's method \cite{Teletzke_1983}.}
\label{fig:1:Ourwork}
\end{center}
\end{figure}

\indent Since both ends of the plug are spherical caps for small
$Ca$, mean curvatures far from the thin film region must match those
of spherical caps. As a result, the following constraint needs to be
satisfied near the spherical cap: $d^2 h/dz^2 = 1$. We may rewrite
this constraint in terms of reduced variables: $\left.(3Ca)^{2/3}\,
(h^*)^{-1}\cdot d^2 \eta/d\xi^2 \right|_{\xi\rightarrow \infty} =
1$. By using asymptotic solutions of $\eta$ given in
Eqn.(\ref{eqn:1:parabolas}), we get:
\begin{equation}
\frac{(3Ca)^{2/3}} {h^*}\cdot P^{F,R} = 1 \, .
\label{eqn:1:Pconstraint}
\end{equation}
\noindent Since $h^*$ is fixed for a given plug, the above equation
implies that $P^F = P^R$. So we can drop the superscript from now
on. For a semi-infinite plug, we shall use Bretherton's result for
$h_\infty$. This leads to the limit value of $P$ as the plug length
goes to infinity: $P_{\infty} = 0.643 $. As derived in detail in the
Appendix of this paper, the dependence of the curvature $P$ on the
plug length $\xi_p$ for finite long plugs is given by: $P(\xi_p) =
P_\infty -1.38 \,\delta_2(\xi_p) + 0.48 \,\delta_3(\xi_p)$, where
functions $\delta_2$ and $\delta_3$ are linear combinations of modes
$\psi_2$ and $\psi_3$, defined in Eqn.(\ref{eqn:1:delta23exp}). By
using the constant curvature constraint derived above, we have:
\begin{eqnarray}
\frac{h^*}{h_{\infty}} &=& \frac{P(\xi_p)}{P_{\infty}} \nonumber \\
&=& 1 - 2.15 \, \delta_2(\xi_p) + 0.75 \, \delta_3(\xi_p) \, .
\label{eqn:1:ourh*}
\end{eqnarray}
\noindent This is the dependence of $h^*$ on the dimensionless plug
length $\xi_p$. It can be easily transformed to represent the
dependence on $L_p$ via the scaling relation: $L_p = \xi_p \,
P(\xi_p) \cdot (3Ca)^{1/3}$. Thus the thickness $h^*$ oscillates
around $h_\infty$ with an amplitude that decays as $L_p$ increases,
and with a decay length of order $L$. Strictly speaking, the above
derivation is only valid in the regime: $\xi_p \gg 1$, where
$\delta_2$ and $\delta_3$ are arbitrarily small. For illustrative
purpose, we extrapolate our analytic results to Teletzke's
short-plug regime and make a comparison between data obtained via
two methods in Fig.\ref{fig:1:Ourwork}. Deviation between two sets
of data is obvious for short plugs, where Teletzke's method is more
accurate. For short plugs in the Teletzke's regime, contributions
from all modes are important. Different modes interact nonlinearly,
which leads to a much stronger finite-length effect, as shown in
Fig.\ref{fig:1:Teletzke}. Around $L_p \sim 6 L_\infty$ the data
obtained via Teletzke's method begin to merge with our curve. This
is the regime where the prediction in Eqn.(\ref{eqn:1:ourh*})
becomes valid. \\

\indent In this section, we completed the previously unstudied part
of the finite plug speed problem. Together with Teletzke's method,
we can now predict how fast plugs of different length will move in
smooth cylindrical tubes. Evidently the speed dependence on the plug
length is governed by the solution modes $\psi_{1,2,3}$ in the thin
film region. For long plugs studied in this section, their speeds
are the most sensitive to the mode with the longest wavelength:
$\lambda_{max} \equiv max (|Re(\nu_i^{-1})|)$. In smooth tubes, this
$\lambda_{max}$ is much smaller than the tube radius $R$ for small
capillary numbers of interest. The asymptotic corrections of the
finite plug speed is exponentially weak, with a decay length
$\lambda_{max}$. For very long plugs with $L_p \gg \lambda_{max}$,
the speed is almost the same as that of the semi-infinite one. As
discussed in the next section, tube roughness can modify the
spectrum of solution modes in the thin film region dramatically so
that $\lambda_{max}$ may become comparable to or larger than $R$.
This is crucial in understanding some interesting experimental
phenomena involving the relative motion between finite plugs of
different length in non-smooth capillary tubes. \\

\section{Relaxation modes in the thin film region with tube roughness}

\begin{figure}
\begin{center}
\includegraphics[angle=-90, width=4.8in]{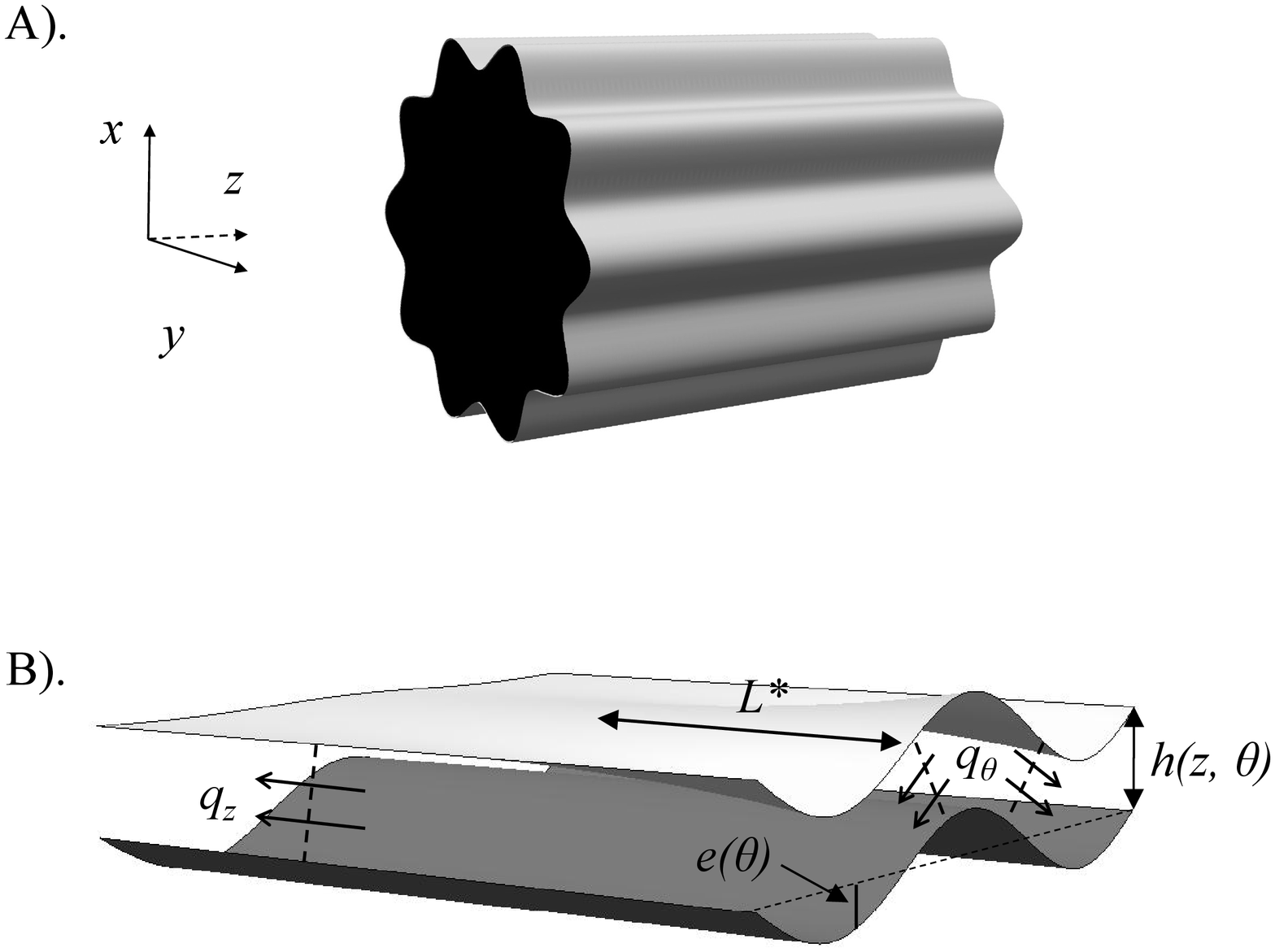}
\caption{A). A sketch of a capillary tube with azimuthal roughness.
B). A plug in the rough tube showing the quantities $e(\theta)$,
$h(z,\theta)$, $q_\theta$, $L^*$ defined in the text.}
\label{fig:2:Roughness}
\end{center}
\end{figure}

\indent In this section, we investigate the shape of the thin film
region of semi-infinite plugs in tubes with roughness. Here tube
roughness is defined as surface deformation of the tube wall, whose
shape is not cylindrical. For simplicity, the only type of roughness
that we will consider in the following discussion is azimuthal
roughness. A tube with the azimuthal roughness preserves its
translation symmetry in the axial direction, but cross sections are
no longer circular. A sketch of such a tube is plotted in
Fig.\ref{fig:2:Roughness}(A). The cross section of the rough tube is
parameterized in the polar coordinate system as: $\tilde{R}(\theta)
= R + e(\theta)$, where $R$ is the smooth tube radius and $\theta$
is the polar angle. The roughness function $e(\theta)$ measures the
deformation. In order to isolate the regime of small roughness we
may express $e(\theta)$ as: $e(\theta) = \epsilon \,
\alpha(\theta)$. Here $\epsilon$ is some small parameter much less
than the Bretherton's thin film thickness $h_\infty$, and
$\alpha(\theta)$ is a function of order unity. The tube roughness
breaks rotational symmetry in the azimuthal plane. As a result, the
plug shape now depends on both the $z$ and the $\theta$ coordinate:
$h(z,\theta)$. Local lubricating film thickness is defined by the
distance between the tube wall and the plug surface: $h(z,
\theta)+e(\theta)$, as shown in Fig.\ref{fig:2:Roughness}(B). As in
the previous section, we nondimensionalize the system by using the
smooth tube radius $R$, the plug speed $U$ and the capillary
pressure $\gamma/R$. As derived later in this section, the
semi-infinite plug shape in the thin film region can be expressed
as:
\begin{equation}
h(z,\theta) \simeq H_\infty + \sum_k C_k \, \psi_k(\theta) \,
\exp{(\nu_k z)} \, \,
\end{equation}
\noindent where $\psi_k$ and $\nu_k$ are eigenfunctions and
eigenvalues governing the relaxation of the plug shape to the
cylindrical asymptote $H_\infty$. Relaxation mode coefficients $C_k$
are determined by boundary conditions away from the thin film
region. In Section 4, we will estimate the order of magnitude of
$C_k$ near the transition between the thin film and the
remobilization region. \\

\subsection{The plug shape equation in rough tubes}

\indent \indent As in the case of the smooth tube, the plug shape is
dictated by the local carrier fluid flux $\vec{q}(z,\theta)$:
\begin{equation}
\vec{q}(z, \theta) = \int_{-e(\theta)}^{h(z,\theta)} \vec{v}(z,
\theta, r) \,dr \, ,
\end{equation}

\noindent where $\vec{v}(z, \theta, r)$ is the carrier fluid
velocity. As shown in Fig.\ref{fig:2:Roughness}(B), $\vec{q}$ has
both an axial component $q_z$ and an azimuthal component $q_\theta$.
In the limit of lubrication approximation, and with the same
boundary conditions as in the smooth case, these fluxes are given by
\cite{deGennes_2003}:
\begin{eqnarray}
q_z &=& - (h + e) - \frac{(h+e)^3}{3Ca}\, \partial_z P \, ,
\nonumber \\
q_\theta &=& - \frac{(h+e)^3}{3Ca}\, \partial_\theta P \, ,
\label{eqn:2:flows}
\end{eqnarray}
\noindent where $P$ is the capillary pressure on the plug surface.
In our reduced units (\cite{Howell_2003}): $P = -h - (\partial_z^2 +
\partial_\theta^2)h$. Eqn.(\ref{eqn:2:flows}) resembles the usual
Poiseuille's law for thin layer fluid fluxes \cite{deGennes_2003}.
Rearranging terms in the equation of $q_z$, we get:
\begin{equation}
\partial_z^3 h= \frac{3Ca}{(h+e)^3}\cdot(q_z+h+e)
-\partial_z \, (1+\partial_\theta^2) h \, .\label{eqn:2:qzrearrange}
\end{equation}
\noindent This is the counterpart of the Teletzke's plug shape
equation, Eqn.(\ref{eqn:1:shapeODE}), in rough tubes. Without
roughness, $q_z$ is a constant and this equation is sufficient to
determine the plug shape $h(z)$. However, $h$ now depends on both
$\theta$ and $z$ coordinates, the flux $q_z$ is no longer constant
and another equation describing the variation of $q_z$ is needed to
determine the plug shape. This additional equation is the
incompressibility constraint: $\partial_z q_z +
\partial_\theta q_\theta = 0$, which leads to the differential equation:
\begin{eqnarray}
\partial_z q_z&=& \partial_\theta \, \frac{(h+e)^3}{3Ca} \,
\partial_\theta P \nonumber \\
&=& - \partial_\theta \, \frac{(h+e)^3}{3Ca} \,
\partial_\theta \,\left[(1+\partial_\theta^2)\, h + \partial_z^2 h\right] \,
. \label{eqn:2:shapegov}
\end{eqnarray}
\noindent Combining Eqn.(\ref{eqn:2:qzrearrange}) and
(\ref{eqn:2:shapegov}), we get a complete set of differential
equations dictating both the plug shape $h$ and the flux $q_z$:
\begin{equation}
\left\{
\begin{array}{lll}
\partial_z h   & =  & h' \\
\partial_z h'  & =  & h'' \\
\partial_z h'' & =  & D^{-1}\cdot(q_z+h+e)
-(1+\partial_\theta^2)\,  h' \\
\partial_z q_z & =  & - \partial_\theta \, D \,
\partial_\theta \, \left[(1+\partial_\theta^2)\, h + h''\right]
\,\,\,\,\,\,\,\,\,\,\, . \label{eqn:2:ders}
\end{array}\right.
\end{equation}
\noindent The first two equations are simply definition equations of
variables $h'$ and $h''$. And the parameter $D$ is given by: $D =
(h+e)^3/(3Ca)$. Evidently, $z$ and $\theta$ derivatives are
separated to two sides of the equation. \\

\indent In the smooth tube the plug surface has rotational symmetry,
so the last equation of Eqn.(\ref{eqn:2:ders}) is simplified as:
$\partial_z q_z = 0$. Obviously, it implies a constant flux $q_z$.
Defining this flux as $-h^*$ and using it in the third equation of
Eqn.(\ref{eqn:2:ders}), we get:
\begin{equation}
\partial_z^3 h = \frac{3Ca \cdot(h - h^*)}{h^3}- \partial_z h \, .
\label{eqn:2:Teletzkewithhp}
\end{equation}
\noindent Finally, we saw in the previous section that in the region
of interest $h$ varies over a length scale of about $L \sim
Ca^{1/3}$. So the slope term $\partial_z h$ in
Eqn.(\ref{eqn:2:Teletzkewithhp}) is negligible compared to the other
two terms for small capillary numbers. And the Teletzke's smooth
plug shape equation, Eqn.(\ref{eqn:1:shapeODE}), is obtained after
dropping $\partial_z h$. \\

\indent Returning to the case of nonzero $e(\theta)$, for
convenience, we define a plug-shape vector: $\vec{\psi}\equiv
\left\{h, \, h', \, h'', \,q_z + e\right\}^T$, where the superscript
$T$ indicates the matrix transpose operation. Then we can rewrite
the above set of differential equations in a concise matrix form:
\begin{eqnarray}
\partial_z \vec{\psi} &=& \left(
\begin{array}{cccc}
0 & 1 & 0 & 0 \\
0 & 0 & 1 & 0 \\
D^{-1} & -(1+\partial_\theta^2) & 0 & D^{-1} \\
-\partial_\theta \,D \,\partial_\theta\,(1+\partial_\theta^2) & 0 &
-\partial_\theta\, D \,\partial_\theta & 0
\end{array}
\right) \, \vec{\psi}  \nonumber \\
&& \nonumber \\
&\equiv& \mathbf{M}(h) \, \vec{\psi} \, . \label{eqn:2:shape}
\end{eqnarray}
\noindent The nonlinearity of this equation arises because of the
dependence of $D$ on $h$. In the following discussion, the roughness
function $e(\theta)$ is assumed to be of a simple sinusoidal form:
\begin{equation}
e(\theta) = e_0 \, \cos{(m\theta)} \, , \label{eqn:2:roughnessform}
\end{equation}

\noindent where $m$ are positive integers indicating different
roughness modes. The roughness amplitude $e_0$ is assumed to be
small in the regime of interest: $e_0 \ll h_\infty$. As discussed in
detail later, Eqn.(\ref{eqn:2:shape}) can be reduced to a linear
form in the thin film region. More general roughness functions can
be expanded in terms of the Fourier series and analyzed
correspondingly. \\

\subsection{Asymptotic shape of semi-infinite plugs}

\indent \indent In this section, we consider the asymptotic shape of
the front region of a semi-infinite plug of arbitrarily long length
$L_p \gg 1$. A similar analysis can easily be applied to the rear
region of the semi-infinite plug. One immediate concern with these
long plugs is that they may be unstable against fission via the
Rayleigh instability \cite{deGennes_2003}. This instability is
independent of the steady state effect that we discuss here. Indeed,
for small capillary numbers, the growth rate of the Rayleigh
instability becomes
negligible \cite{deGennes_2003}.\\

\indent As with the smooth tube case discussed above, we may fix the
origin of $z$ coordinate at the transition between the thin film and
the remobilization region. Far inside the thin film region where $z
\ll -1$, the front semi-infinite plug restores the translation
symmetry in the $z$ direction. It implies vanishing $z$ derivatives
in this limit, \textit{i.e.} $\partial_z \vec{\psi} (\infty) = 0$.
Evidently, $h'(\infty)$ and $h''(\infty)$ are zero. The third
equation of Eqn.(\ref{eqn:2:shape}) leads to the condition: $h =
-q_z - e$. And the asymptotic plug shape $h(\infty)$ is determined
by the last equation of Eqn.(\ref{eqn:2:shape}): $-\partial_\theta
\,D \,\partial_\theta\,(1+\partial_\theta^2) h = 0$. One obvious
solution is that $h(\infty) \equiv H_\infty$, where $H_\infty$ is a
constant. This solution corresponds to a cylindrical plug centered
inside the tube. There are two other solutions: $H_1 \cos\theta$ and
$H_2 \sin\theta$, which correspond to cylindrical plugs shifted
laterally inside the tube. The translation symmetry far inside the
thin film region leads to the cylindrical plug shape with constant
capillary pressure on the plug surface and vanishing flux
$q_\theta$. For convenience, we focus on the centered asymptote. The
implication of non-centered solutions will be addressed in the
Discussion section. \\

\indent The centered solution $H_\infty$ measures the magnitude of
average $q_z$ per unit circumference:
\begin{eqnarray}
\bar{q} &\equiv&\frac{1}{2\pi} \int_0^{2\pi} q_z(\infty) \, d\theta
\nonumber
\\
&=&\frac{1}{2\pi} \int_0^{2\pi} (-H_\infty - e) \, d\theta \nonumber
\\
&=& -H_\infty \, . \label{eqn:2:qbar}
\end{eqnarray}
\noindent So $H_\infty$ has the same physical meaning as $h_\infty$
in the smooth tube case, and dictates the relative motion between
the plug and the carrier fluid: $U/V = 1 + 2H_\infty $. For tubes
with small roughness, the difference between $H_\infty$ and
$h_\infty$ is also very small: $|H_\infty - h_\infty| \ll h_\infty$.
The leading order correction in $H_\infty$ due to the tube roughness
appears in
the second order of $e$. \\

\subsection{Eigenmodes of relaxation in the thin film region}

\indent \indent At finite $z$ values, the semi-infinite plug shape
deviates from its cylindrical asymptote. The plug-shape vector
$\vec{\psi}$ is rewritten in the following form:
\begin{eqnarray}
\vec{\psi} &=& \vec{\psi}(\infty) + \delta \vec{\psi} \nonumber \\
&\equiv& \left\{H_\infty+\delta h, \, \delta h', \, \delta h'', \,
-H_\infty + \delta q_z\right\}^T \, , \label{eqn:2:delMdef}
\end{eqnarray}
\noindent where $\delta \vec{\psi} \equiv \left\{\delta h, \, \delta
h', \, \delta h'', \, \delta q_z \right\}^T $ measures the amount of
shape correction. In the thin film region, $\delta \vec{\psi}$
satisfies the relation: $\left| \, \delta \psi^i \, \right| <<
\left| \, \psi^i (\infty) \, \right|$, where the superscript $i$
indicates different vector components. So the plug shape equation
(\ref{eqn:2:shape}) can be linearized by using the approximation: $D
\simeq D_1 \equiv (H_\infty + e)^3/(3Ca)$, and we get:
\begin{eqnarray}
\partial_z \delta\vec{\psi} &=& \left(
\begin{array}{cccc}
0 & 1 & 0 & 0 \\
0 & 0 & 1 & 0 \\
D_1^{-1} & -(1+\partial_\theta^2) & 0 & D_1^{-1} \\
-\partial_\theta \,D_1 \,\partial_\theta\,(1+\partial_\theta^2) & 0
& -\partial_\theta\, D_1 \,\partial_\theta & 0
\end{array}
\right) \, \delta\vec{\psi} \nonumber \\
&& \nonumber \\
&\equiv& \mathbf{M}_1 \, \delta{ \vec{\psi}} \, ,
\label{eqn:2:shapedeltaM}
\end{eqnarray}
\noindent Evidently, we can solve the above equation via the
technique of separation of variables. The general solution of
$\delta{\vec{\psi}}$ is given by:
\begin{equation}
\delta{\vec{\psi}} = \sum_{k} C_k \, \vec{\psi}_k(\theta) \,
\exp{(\nu_k z)} \, ,
\end{equation}
\noindent where $\vec{\psi}_k(\theta)$ and $\nu_k$ are eigenvectors
and eigenvalues of the matrix operator $\mathbf{M}_1$, with
$\mathbf{M}_1 \, \vec{\psi}_k(\theta) = \nu_k \,
\vec{\psi}_k(\theta)$. Thus in the thin film region,
$\vec{\psi}_k(\theta) \, \exp{(\nu_k z)}$ are the eigenmodes
dictating the relaxation of the plug surface to its cylindrical
asymptote. The speed of relaxation is determined by the decay length
of each mode $\lambda_k$, defined as $\lambda_k \equiv |Re(1/
\nu_k)|$. The larger the value of $\lambda_k$, the slower of the
relaxation process. Relaxation modes with positive $Re(1/\nu_k)$ are
dominant for the front semi-infinite plug. Likewise, negative
ones are dominant for the rear semi-infinite plug. \\
\begin{figure}
\begin{center}
\includegraphics[angle=-90, width=4.6in]{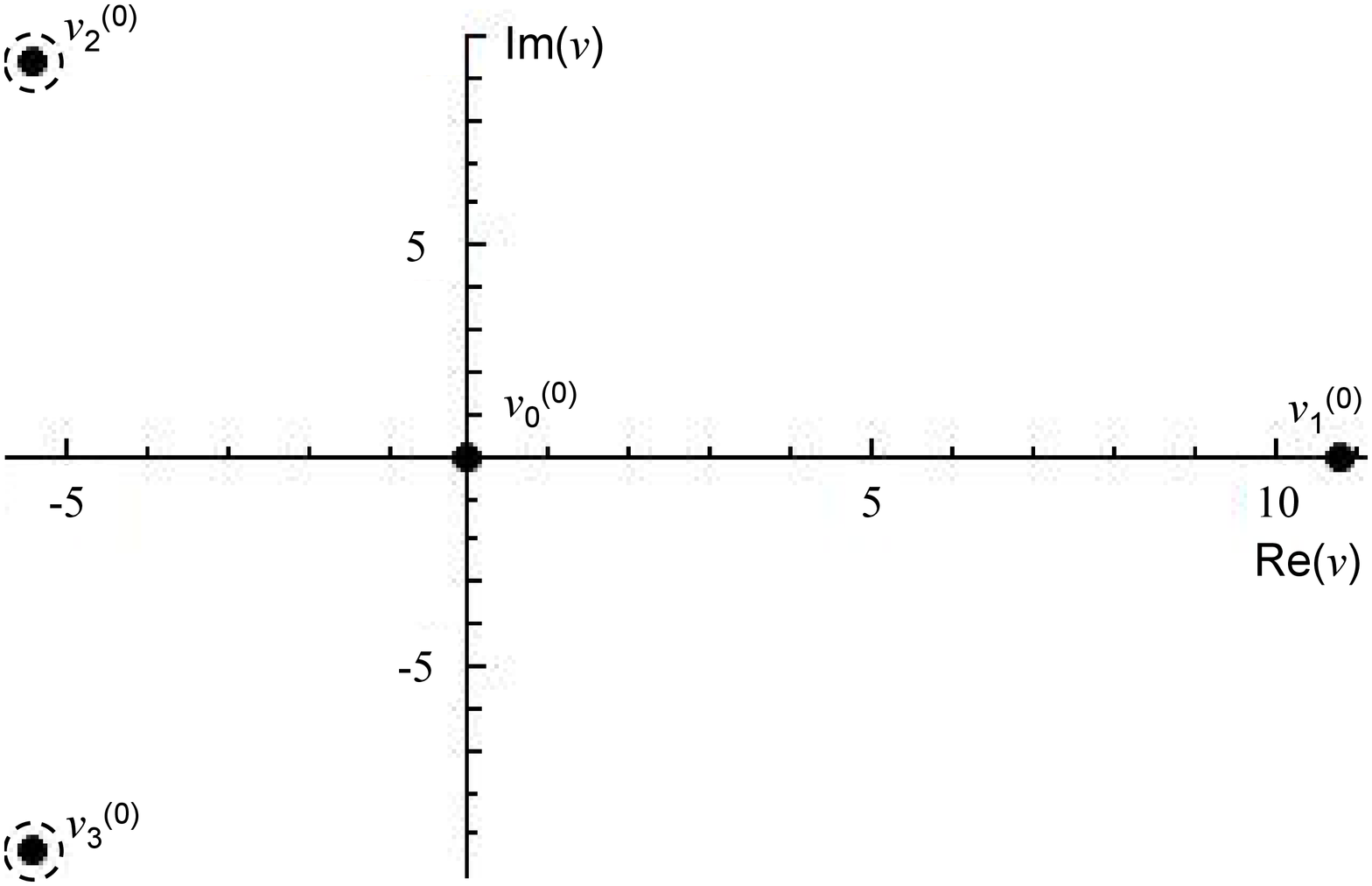}
\caption{Zeroth harmonic spectrum of the semi-infinite plug shape in
the thin film region for tubes with small sinusoidal roughness.
Capillary number's value: $Ca = 10^{-3}$.}
\label{fig:2:smoothspectrum}
\end{center}
\end{figure}

\indent In the regime of vanishingly small roughness $e(\theta)$,
Eqn.(\ref{eqn:2:shapedeltaM}) can be simplified further as follows:
\begin{eqnarray}
\partial_z \delta\vec{\psi} &=& \left(
\begin{array}{cccc}
0 & 1 & 0 & 0 \\
0 & 0 & 1 & 0 \\
D_0^{-1} & -(1+\partial_\theta^2) & 0 & D_0^{-1} \\
-D_0\,(\partial_\theta^2+\partial_\theta^4) & 0 &
-D_0\,\partial_\theta^2 & 0
\end{array}
\right) \, \delta\vec{\psi} \nonumber \\
&& \nonumber \\
&=& \mathbf{M}_0 \, \delta{ \vec{\psi}} \, ,
\label{eqn:2:simshapedeltaM}
\end{eqnarray}
\noindent where the coefficient $D_0$ is now given by: $D_0 \equiv
h_\infty^3 /(3Ca)$. Evidently the matrix operator $\mathbf{M}_0$ has
no explicit dependence on the $\theta$ coordinate, so its
eigenvectors correspond to different harmonics in the Fourier series
of $\theta$. Due to the form of the roughness function $e(\theta)$
chosen, we may focus on cosine series solutions with the same
symmetry: $\sum C_n \cos{(n m \theta)}$. Here $n$ is the harmonic
number. Up to the first order in the roughness $e$, we need to
analyze the zeroth and first harmonics, with each harmonic leading
to four eigenmodes:
\begin{eqnarray}
\delta \vec{\psi} &\simeq& \delta \vec{\psi}^{(0)} + \delta
\vec{\psi}^{(1)}
\nonumber \\
&=& \sum_{k=0}^3 C_k^{(0)}\,\vec{\psi}_k^{(0)}
\,\exp{(\nu_k^{(0)}z)} +\sum_{k=1}^4 C_k^{(1)}
\,\vec{\psi}_k^{(1)}\cos{(m\theta)}\,\exp{(\nu_k^{(1)} z)} \, ,
\end{eqnarray}
\noindent where superscripts (0) and (1) stand for the zeroth and
the first harmonics respectively. The zeroth harmonic solution
$\delta \vec{\psi}^{(0)}$ preserves the rotational symmetry and
satisfies the equation:
\begin{equation}
\partial_z \delta\vec{\psi}^{(0)} = \left(
\begin{array}{cccc}
0 & 1 & 0 & 0 \\
0 & 0 & 1 & 0 \\
D_0^{-1} & -1 & 0 & D_0^{-1} \\
0 & 0 & 0 & 0
\end{array}
\right) \, \delta\vec{\psi}^{(0)} \, .
\end{equation}
\noindent The above matrix operator is a four-by-four matrix and
does not depend on the form of the roughness function. So the zeroth
harmonic spectra of different roughness functions are the same. For
small capillary numbers, four eigenvalues of this matrix
$\nu_k^{(0)}$ are given by: $\nu_0^{(0)} = 0$, $\nu_1^{(0)} =
D_0^{-1/3}$, $\nu_2^{(0)} = D_0^{-1/3}(-1/2 + i \sqrt{3}/2)$, and
$\nu_3^{(0)} = D_0^{-1/3}(-1/2 - i \sqrt{3}/2)$. There is no plug
surface relaxation associated with the mode $\nu_0^{(0)}$. It
corresponds to the degree of freedom to fix the radius of the
cylindrical asymptote as $z \rightarrow \infty$. The other three
modes $\nu_1^{(0)}$, $\nu_2^{(0)}$ and $\nu_3^{(0)}$ correspond to
the ones found in the previous section for the smooth tube, up to a
difference in the choice of units. For illustrative purpose, the
zeroth harmonic spectrum is plotted in
Fig.\ref{fig:2:smoothspectrum}. Obviously, $\nu_2^{(0)}$ and
$\nu_3^{(0)}$ are the relaxation modes with the longest decay length
$\lambda_{max}^{(0)}$ in this spectrum. \\
\begin{figure}
\begin{center}
\includegraphics[angle=-90, width=4.6in]{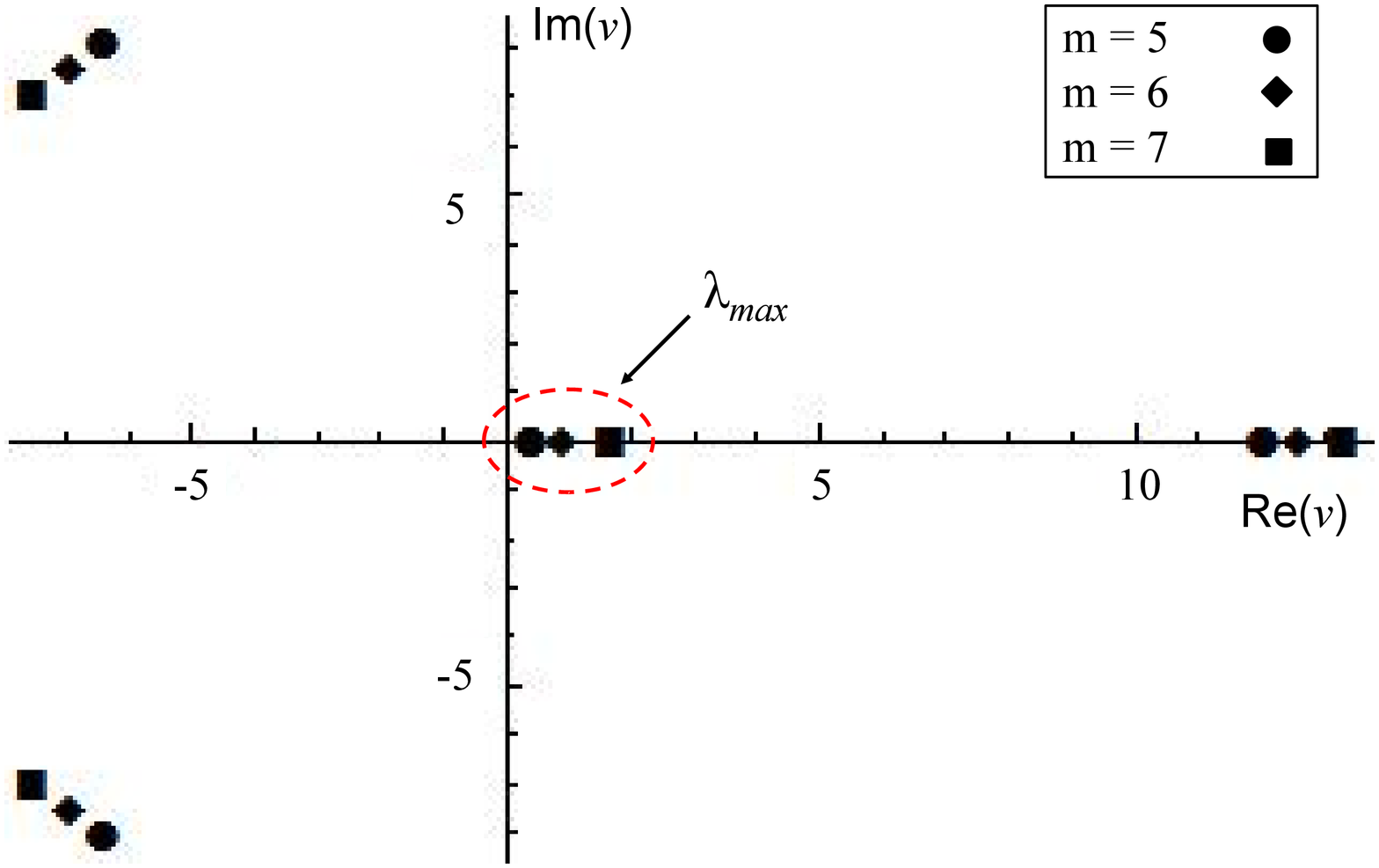}
\caption{First harmonic spectrum of the semi-infinite plug shape in
the thin film region for tubes with small sinusoidal roughness.
Cases with $m =$5, 6, 7 are plotted. Capillary number's value: $Ca =
10^{-3}$.} \label{fig:2:roughspectrum}
\end{center}
\end{figure}

\indent A major difference between the rough tube and the smooth one
comes from the first harmonic solution $\delta \vec{\psi}^{(1)}$. It
satisfies the equation:
\begin{equation}
\partial_z \delta\vec{\psi}^{(1)} = \left(
\begin{array}{cccc}
0 & 1 & 0 & 0 \\
0 & 0 & 1 & 0 \\
D_0^{-1} & -1+m^2 & 0 & D_0^{-1} \\
D_0\,(m^2 - m^4) & 0 & D_0\, m^2 & 0
\end{array}
\right) \, \delta\vec{\psi}^{(1)} \, .
\end{equation}
\noindent Four eigenvalues are solutions of the characteristic
equation:
\begin{equation}
\nu^4 +(1-2m^2)\, \nu^2 - D_0^{-1} \, \nu + m^4 - m^2 = 0 \,
.\label{eqn:2:character}
\end{equation}
\noindent The first harmonic spectra with the roughness mode number
$m = $5, 6, and 7 are plotted in Fig.\ref{fig:2:roughspectrum}.
Compared to the zeroth harmonic spectrum shown in
Fig.\ref{fig:2:smoothspectrum}, the mode with the longest decay
length $\lambda_{max}^{(1)}$ now lies near the origin and has much
longer decay length. For small values of $m$, \textit{i.e.} long
wavelength roughness, $\lambda_{max}^{(1)}$ becomes comparable to or
even larger than the tube radius $R$. Obviously,
$\lambda_{max}^{(1)} > \lambda_{max}^{(0)}$ in the regime of
interest. So $\lambda_{max}^{(1)}$ is the new slowest decay mode in
rough tubes after analyzing both the zeroth and the first harmonic
spectra. So we will drop the superscript $(1)$ and refer to it as
$\lambda_{max}$ from now on. \\

\indent To better understand the origin of these slow relaxation
modes in tubes with roughness, we propose a following physical
explanation of $\lambda_{max}$. The remobilization region ends at
about distance $z\sim L_\infty$ from the spherical cap. At this
transition between the remobilization region and the thin film
region, the plug shape has not reached its cylindrical asymptote
yet. We assume that the plug surface has a typical height variation
of about $\Delta h$. A sketch of such a plug cross section is given
in Fig.\ref{fig:2:Roughness}(B). Non-cylindrical plug shape leads to
unbalanced capillary pressure $P$ in the azimuthal plane and a
non-zero lateral flow $q_\theta$. In the thin film region, the plug
surface relaxes toward the cylindrical asymptote via this flow
$q_\theta$. Suppose that the relaxation length scale is given by
$L^*$. It can be estimated as the product between the carrier fluid
speed $U$ in this region and a relaxation time scale $T$ over which
$q_\theta$ smoothes the plug surface. In the following analysis, we
will estimate the length scale $L^*$ and obtain its scalings with
other system parameters. For this part of discussion only, we will
work in the lab units. \\

\indent The expression of the lateral flux $q_\theta$ is given by
Eqn.(\ref{eqn:2:flows}). In the lab units, we have:
\begin{equation}
q_\theta = - \frac{(h+e)^3}{3\mu R}\, \partial_\theta P \, .
\label{eqn:2:qtheta}
\end{equation}
\noindent As mentioned above, the capillary pressure $P$ is no
longer a constant in the azimuthal plane for the non-cylindrical
plug shape. The length scale in the $\theta$ direction is defined by
the roughness wavelength $l_\theta \sim R / m$. So we may estimate
$\theta$ derivatives as: $R^{-1} \, \partial_\theta h\sim (m/R)
\cdot \Delta h $. Since the $z$-direction plug surface adjustment
mostly happens in the remobilization region, we expect that $\theta$
derivatives are typically much larger than $z$ derivatives in the
thin film region. The limit of this assumption will be derived
later. So the capillary pressure $P$ is approximated as follows:
\begin{eqnarray}
P  &=& -\gamma\, \left[ \, \partial_z^2 h - \,
\frac{(1+\partial_\theta^2)h}{R^2}\right] \nonumber \\ &\sim& \gamma
\, \frac{\Delta h}{(R/m)^2} \, . \label{eqn:3:curvatures}
\end{eqnarray}
\noindent By definition, the surface relaxation time scale $T$ is
given by:
\begin{equation}
T \simeq \frac{\Delta h}{R^{-1}\,\partial_\theta q_\theta} \sim
\frac{\mu\,(R/m)^4}{\gamma\, h_\infty^3} \, .
\end{equation}
\noindent The relaxation length scale $L^*$ is estimated as the
product between $U$ and $T$: $L^* \sim Ca\, R^4/(h_\infty^3\, m^4)$.
We may simplify this estimation of $L^*$ further by using the
scaling of $h_\infty$ and get:
\begin{equation}
L^* \sim \frac{R}{Ca \cdot m^4} \, . \label{eqn:2:L^*2}
\end{equation}
\noindent For a self-consistent derivation, $m$'s value is
constrained by the approximation used in
Eqn.(\ref{eqn:3:curvatures}). This is equivalent to the condition:
$R^{-1} \,\partial_\theta h \gg \partial_z h$, which yields the
relation: $l_\theta \ll L^*$. By using the estimation of $L^*$
obtained above, we get the valid regime of the roughness mode
number: $m \ll Ca^{-1/3}$. It corresponds to capillary tubes with
long wavelength roughness. Obviously $L^*$ is much longer than the
smooth remobilization length $L_\infty$ in this regime of
roughness.\\

\indent The length scale $L^*$ is actually the longest decay length
$\lambda_{max}$ in the eigenvalue spectrum derived previously. It
corresponds to the smallest solution of Eqn.(\ref{eqn:2:character}),
which goes to zero as $Ca$ approaches zero. For this root, the
$D_0^{-1}\, \nu$ term dominates the higher power terms of $\nu$, and
Eqn.(\ref{eqn:2:character}) simplifies as follows:
\begin{equation}
-D_0^{-1} \cdot \lambda_{max}^{-1} + m^4 - m^2 = 0 \, .
\end{equation}
\noindent For an order of magnitude estimation, we may drop the term
$m^2$ relative to $m^4$ and get: $\lambda_{max} \sim Ca
/(h_\infty^3\, m^4)$. This is the same scaling relation as that of
$L^*$, up to a difference in the choice of units. Thus $L^*$ and
$\lambda_{max}$ measure the same length scale in rough tubes. The
scaling argument presented above physically explains the origin of
large $\lambda_{max}$ values in the rough tube. This new length
scale reflects the slow relaxation of the plug shape via a
vanishingly small lateral fluid flux $q_\theta$. Historically,
people has observed some evidence of this new length scale in
non-cylindrical tubes \cite{Wong_1995}.\\

\section{Amplitude of the plug surface bumpiness in the remobilization
region}

\indent In the previous section, we analyzed the semi-infinite plug
shape in the thin film region and found that the plug surface decays
to its cylindrical asymptote via different relaxation modes. To
proceed further, we must understand how the rough tube wall imposes
a non-cylindrical plug shape in the first place. For small
roughness, the plug shape is dominated by the zeroth and the first
harmonic solutions:
\begin{equation}
h(z,\theta) \simeq H_\infty + \sum_{k=1}^3 C_k^{(0)}
\,\exp{(\nu_k^{(0)}z)} +\sum_{k=1}^4 C_k^{(1)}
\,\cos{(m\theta)}\,\exp{(\nu_k^{(1)} z)} \, , \label{eqn:4:modes}
\end{equation}
\noindent where the coefficients $C_k^{(0)}$ and $C_k^{(1)}$ are
determined from the boundary conditions away from the thin film
region. In this section, we will treat the front and the rear
semi-infinite plugs simultaneously, with the understanding that only
the modes with the correct sign of $Re(1/\nu_k)$ are excited for a
single semi-infinite plug. The $z$ coordinate origin is fixed at the
transition between the thin film and the remobilization region as in
Section 2. In that section we have shown that the zeroth harmonic
coefficients $C_k^{(0)}$ are of the order of $h_\infty$. We will see
later in this section that the first harmonic coefficients
$C_k^{(1)}$ are of the order of $e_0$. The exact numerical solutions
will be presented in our future work. \\

\indent In the remobilization region, the governing equation of the
plug shape $h(z,\theta)$ is derived by combining the last two
equations in Eqn.(\ref{eqn:2:ders}):
\begin{equation}
\partial_z(h+e) -
\frac{1}{3Ca}\left[\partial_z(h+e)^3\partial_z +
\partial_\theta(h+e)^3\partial_\theta\right] (1+\partial_z^2+\partial_\theta^2)h = 0 \,
.\label{eqn:4:hfull}
\end{equation}
\noindent This equation can be simplified further by noticing the
fact that in the remobilization region, $h$ varies over about
$(3Ca)^{1/3}$ in the $z$ direction and about $m^{-1}$ in the
$\theta$ direction. For long wavelength roughness with $m \ll
(3Ca)^{-1/3}$, the following estimation of derivatives is valid :
$\partial_z h \gg \partial_\theta h$. In other words, in the
remobilization region the plug shape $h(z,\theta)$ varies much more
rapidly in the $z$ direction than in the $\theta$ direction. So we
may keep only $z$ derivatives in Eqn.(\ref{eqn:4:hfull}), and get:
\begin{equation}
\partial_z h =
\frac{1}{3Ca} \left[
\partial_z(h+e)^3\partial_z^3\,\right] h \, .
\label{eqn:4:hsimplified}
\end{equation}
\noindent Bretherton's smooth tube solution $h_0(z)$ satisfies the
above differential equation with $e = 0$. In the regime of small
roughness, Eqn.(\ref{eqn:4:hsimplified}) can be solved
perturbatively by expanding around the smooth tube solution
$h_0(z)$: $h(z,\theta) = h_0(z) + u^{(0)}(z) + u^{(1)}(z)\,
\cos(m\theta)$. Here $u^{(0)}$ and $u^{(1)}$ are the leading order
plug shape correction attributed to the tube roughness. For
vanishingly small roughness, the relation $|u^{(0,1)}/h_0| \ll 1$
holds for arbitrary $z$. Since we are interested in the coefficients
of the first harmonic solution, we will focus on the shape
correction $u^{(1)}$ for the rest of this section. Keeping only the
lowest order approximation, $u^{(1)}$ satisfies the following
equation:
\begin{equation}
\left(\partial_z h_0^3 \, \partial_z^3\right) u^{(1)} + (3h_0^2\,
\partial_z^3h_0 - 3Ca) \,
\partial_z u^{(1)} + \left[\partial_z(3h_0^2\, \partial_z^3h_0)\right]
(u^{(1)}+e_0)  = 0\, . \label{eqn:4:h1}
\end{equation}
\noindent This is an inhomogeneous differential equation. The
correct boundary conditions of $u^{(1)}$ are that it goes to zero as
$z$ approaches the spherical cap, and matches the relaxation modes
in Eqn.(\ref{eqn:4:modes}) as $z$ approaches the thin film region.
The general solution of Eqn.(\ref{eqn:4:h1}) takes the form:
$u^{(1)}(z) = e_0 \left(f(z) - 1\right)$. Here $f(z)$ is a
dimensionless function satisfying the homogeneous equation and goes
to 1 as $z$ approaches the spherical cap. By using the rescalings as
in the smooth tube: $h_0 = h_\infty \cdot \eta_0$, and $z = L_\infty
\cdot \xi$, we get the following equation of $f$:
\begin{equation}
\left(\partial_\xi \eta_0^3 \, \partial_\xi^3\right) f +
(3\eta_0^2\,
\partial_\xi^3\eta_0 - 3) \,
\partial_\xi f + \left[\partial_\xi(3\eta_0^2\, \partial_\xi^3\eta_0)\right]
f  = 0\, . \label{eqn:4:feq}
\end{equation}
\noindent After the rescaling both $\eta_0$ and its $\xi$
derivatives are of the order of unity in the remobilization region,
so all coefficients in Eqn.(\ref{eqn:4:feq}) are of the order of
unity. Together with the boundary condition of $f$ near the
spherical cap, the solution $f(z)$ must be of the order of unity in
the remobilization region. By matching with relaxation modes in the
thin film region, we get an order of magnitude estimation of the
first harmonic coefficients: $C_k^{(1)} \simeq e_0$. Thus the
semi-infinite plug shape in the thin film region can be rewritten as
follows:
\begin{equation}
h(z,\theta) \simeq H_\infty\,\left(1 + \sum_{k=1}^3
\tilde{C}_k^{(0)} \,\exp{(\nu_k^{(0)}z)}\right) +\sum_{k=1}^4
\tilde{C}_k^{(1)} \,e(\theta) \,\exp{(\nu_k^{(1)} z)} \, ,
\label{eqn:4:roughshape}
\end{equation}
\noindent where all coefficients $\tilde{C}_k^{(0)}$ and
$\tilde{C}_k^{(1)}$ are of the order of unity. The above estimation
is valid for both the front and the rear semi-infinite plug. \\

\section{Finite plug motion in rough tubes}

\indent  In this section, we qualitatively study the dependence of
the finite plug motion on the plug length in the rough tube,
following the logic of Section 2. For the finite plugs, relaxation
modes of both signs of $Re(1/\nu_k)$ are excited in the thin film
region. Far inside the thin film region where the plug shape is
almost cylindrical, all the other relaxation modes are vanishingly
small compared to the slowest decay mode $\lambda_{max}$. Thus we
may focus on the effect of $\lambda_{max}$. By defining a
$\tilde{z}$ coordinate whose origin is at some position far inside
the thin film region, the plug shape is approximated as follows:
\begin{equation}
h(z,\theta) \simeq H_\infty + \tilde{C} \,e(\theta)
\,\exp{\left[(\tilde{z}- \ell) /\lambda_{max}\right]} \, ,
\label{eqn:5:hlambda}
\end{equation}
\noindent where $\tilde{z} \equiv z + \ell$. The coefficient
$\tilde{C}$ is of the order of unity as derived in the previous
section. Here the variable $\ell$ measures the half length of the
finite plug. For plugs used in microfluidic experiments, $\ell$ is
usually comparable to or larger than the tube radius $R$
\cite{Ying_2008}. As mentioned previously, the relative motion
between the plug and the carrier fluid is governed by the azimuthal
average of $q_z$:
\begin{eqnarray}
U/V &=& 1 - \frac{1}{\pi} \int_0^{2\pi} q_z \, d\theta \nonumber \\
&=& 1 - 2\bar{q} \, .\label{eqn:3:UVbarq}
\end{eqnarray}
\noindent Far inside the thin film region, the plug shape is almost
cylindrical, which leads to the following estimation of the carrier
fluid fluxes: $q_z \simeq - h - e$, and $q_\theta \simeq 0$. By
using the approximate plug shape in Eqn.(\ref{eqn:5:hlambda}), we
have: $q_z(z,\theta) \simeq -H_\infty - \tilde{C} \,e(\theta)
\,\exp{\left[(z- \ell) /\lambda_{max}\right]}-e(\theta)$. Evidently,
the average flux $\bar{q}$ vanishes in this linear approximation.
Thus any effect on $\bar{q}$ appears in the second or higher order
of the roughness: $\bar{q} + H_\infty = \tilde{B} \,e_0^2 \,\exp(- 2
\ell / \lambda_{max}) + O(e_0^4)$. The asymptotic dependence of the
plug speed on the plug length is exponentially small for long plugs,
and we have:
\begin{equation}
U/V \simeq 1 + 2H_\infty\, \left[1 + \frac{\tilde{B}
\,e_0^2}{H_\infty}\,\exp(- 2 \ell / \lambda_{max})\right] \, .
\label{eqn:3:UVAsymptotic}
\end{equation}
\noindent The above form of length dependence implies an interesting
resonance effect in the rough tube. To illustrate this effect, we
may estimate the speed difference between two plugs with length
$\ell_1$ and $\ell_2$. We also assume that the length difference
between these two plugs, defined as $\Delta \ell \equiv \ell_2 -
\ell_1$, is small such that $|\Delta \ell / \ell_1 | \ll 1$. By
using Eqn.(\ref{eqn:3:UVAsymptotic}), we derive their speed
difference as follows:
\begin{eqnarray}
(U_2 - U_1)/ V &\simeq& 2 \tilde{B} \, e_0^2 \, \left[\exp(- 2\ell_2
/ \lambda_{max})- \exp(- 2\ell_1 / \lambda_{max})\right]
\nonumber\\
&\simeq& 2\tilde{B} \, e_0^2 \,\left(\frac{-\Delta
\ell}{\ell_1}\right) \cdot \left[\exp(- 2\ell_1 /
\lambda_{max})\cdot (2\ell_1/ \lambda_{max})\right] \, ,
\label{eqn:3:U1U2}
\end{eqnarray}
\noindent The function of $\lambda_{max}$ in the last square bracket
above is not monotonic. It has a maximum at $\lambda_{max} =
2\ell_1$. For plugs used in microfluidics measurements, we have:
$\ell_1 \simeq R$. Thus the speed difference of these plugs is
maximized in rough tubes with $\lambda_{max} \simeq 2R$. By using
the scaling of $\lambda_{max}$ derived in Section 3, we get the
roughness mode of resonance: $m_r \sim (Ca)^{-1/4} $. \\

\indent In the rest of this section, we will use an example to
illustrate how the tube roughness enhances the sensitivity of the
plug motion to the plug length. Suppose that we have two finite
plugs of length: $\ell_1 = 0.5R$ and $\ell_2 = R$. Strictly
speaking, Eqn.(\ref{eqn:3:UVAsymptotic}) is valid for the regime of
vanishingly small roughness. However, we expect this estimation to
be at least qualitatively correct for tubes with large roughness
Thus for computation simplicity, we may set the coefficient
$\tilde{B} \, e_0^2 / H_\infty$ to be 1. In the smooth tube case,
$H_\infty = h_\infty$ and $\lambda_{max} = 2L_\infty$. And we get
the speed difference between these two plugs less than $0.01\%$ with
$Ca = 0.001$. Thus the finite plug length effect is negligible in
the smooth tube. On the other hand for tubes with long wavelength
roughness, as shown in Fig.\ref{fig:2:smoothspectrum}, we have
parameter values: $H_\infty \simeq h_\infty$ and $\lambda_{max}
\simeq R$. The plug speed difference can be up to a few percent,
which is significantly larger than that in the smooth tube. \\

\section{Discussion}

\indent In the previous sections we have argued that roughness in a
microchannel can qualitatively alter capillary motion in the
channel. We demonstrated the new effects in the context of an air
plug being pushed through the channel. Roughness creates new modes
of relaxation of the fluid interface. These modes make major changes
in the Bretherton-Teletzke mechanism that sets the film thickness
and governs the plug's speed. In particular, they alter the nature
of the the remobilization region that dictates the thickness of the
liquid film between the plug and the wall. These modes can thus
alter the effect of other variations in the system, such as the
addition of surfactants or the replacement of the air
plug by a viscous fluid. \\

\indent Our focus above has been to consider the effect of plug
length on its speed. This basic effect has practical consequences as
noted in the Introduction. Microfluidic devices for chemical
processing create long trains of droplets that need to remain
separated over long distances. However in practice these droplets
can have slight differences of speed. This can make droplets merge,
thus disrupting the desired behavior. Since the length of these
droplets is not precisely controlled, the length differences could
in principle be responsible for the undesired speed differences.
However, such a length effect is incompatible with the classic
Bretherton explanation for the speed. We showed above that roughness
changes this conclusion qualitatively. \\

\indent This effect complements previously-considered effects of
channel shape. It differs from the roughness effect identified by
Homsy et al. \cite{Krechetnikov_2005}, in which roughness alters the
effective boundary condition at the wall. We showed here that
roughness can also have long-range effects on the thin film
behavior. These effects do not require the qualitative changes in
channel shape found, \textit{e.g.}, for square channels
(\cite{Wong_1995, Wong_1995_2}). Remarkably, even high-wavenumber
roughness can alter the long-range effects if the capillary number
is sufficiently small, as shown in Section 5.\\

\indent Roughness may be viewed as a relevant perturbation to the
plug viewed as a dynamical system. The fast relaxation to the thin
film within a narrow remobilization region seen in a Bretherton plug
depends crucially on its assumed azimuthal symmetry. When this
symmetry is broken by roughness, the relaxation takes on a
qualitatively new behavior. In this sense the symmetric plug is
unstable against the generic un-symmetric behavior discussed above.
For this reason it appears important to consider roughness when
estimating any new aspect of plug behavior. \\

\indent Though our work suggests that roughness is important, our
quantitative exploration of this importance has been very limited.
First, we considered only azimuthal roughness, ignoring any $z$
dependence of the wall position. We believe the nonlocal effects
identified above are not qualitatively altered for generic
roughness, we have not addressed this issue explicitly. We consider
only perturbative effects of roughness in lowest order. Thus we have
no systematic calculation for the speed, which requires the next
order of approximation. Our analysis of the remobilization region
was limited to a scaling discussion, with no systematic results. We
didn't carry out more detailed calculations because the perturbation
approach itself has serious flaws. It becomes qualitatively
inadequate for low capillary number, where the unperturbed film
thickness becomes smaller than any roughness. In this regime the
interface is mainly supported by asperities where the film thickness
is much smaller than its average. In order to get quantitative
information on how roughness and length affect speed, we plan a
numerical integration of Eqn.(\ref{eqn:2:shape}). \\

\indent Our analysis unearthed a peculiar behavior for the
lowest-wavenumber perturbation, namely $m=1$. This mode amounts to a
lateral displacement of the plug in a circular channel. Remarkably
this perturbation does not relax since it does not alter the
circular shape of the plug. Centering effects of particles in tubes
have been discussed \cite{Brenner_1963}, but we are not aware of an
analysis suited to Bretherton plugs. \\

\indent The roughness effects shown here suggest a new regime of
interest in microfluidic capillary flow. In this rough-channel
regime, the symmetry of the Bretherton plugs is lost, but the fluid
coating the interface is still a thin film in lubrication flow. Thus
these channels are distinct from, e.g., the square channels
discussed by Wong (\cite{Wong_1995}), where the lubrication
approximation is inadequate. In addition to capillary number and the
plug aspect ratio, two dimensionless variables characterize a system
in this regime. The first is the roughness wavelength $R/m$ relative
to the Bretherton film thickness $h_\infty$. The second is the
amplitude $e_0$ relative to this thickness. We have noted above the
special effects to be explored when the decay length is comparable
to the plug length. We have also suggested an asperity-dominated
regime to be expected when the dimensionless amplitude becomes
large. Naturally the chief experimental impact of these effects
comes when the plug is replaced by a droplet of viscous liquid. The
effects anticipated above should all have counterparts for such
liquid droplets. \\

\indent Our predictions, though qualitative, are subject to
experimental tests. An obvious test is to create tubes with
controlled corrugated profiles and verify that when the roughness
amplitude is comparable to the predicted Bretherton film thickness,
the speed increment $U - V$ changes significantly relative to that
of a smooth tube with the same cross section. One can also create
corrugations with a wavelength where sensitivity to plug length is
expected and look for significant variations in $U - V$ when the
roughness amplitude is comparable to the Bretherton thickness. Such
tests could be done at the microfluidic scale or at
larger scales. \\

\indent More primitive observations using microfluidic channels are
also suggested by our analysis. One may readily prepare circular
tubes with 200 micron cross section and controllable flows with
capillary numbers of order $10^{-5}$ \cite{Ying_2008}. For such
flows, the predicted Bretherton thickness is 60 nanometers, so that
roughness amplitudes of this order should be significant. AFM
measurements on these tubes show roughness with $m$ of order $10^2$
and amplitude of order 50 nanometers. This amplitude is evidently
large on the scale of interest. Since the roughness varies randomly
along the tube, one expects statistical fluctuations in the speed
$U$ along the tube. Some of these fluctuations can be due to
differences in cross sectional area, so that they cause the fluid
speed $V$ to change with position. In addition one expects
variations in $U - V$ owing to our mechanism. The indication that
these variations are caused by roughness is that $U - V$ is
determined by position. Thus \textit{e.g.} large $U - V$ values
occur at particular places along the tube. \\

\section{Conclusion}

\indent The subtle interplay of nonlocal forces that controls a
microfluidic droplet is a classic example of the power of simple
capillary flow to produce nontrivial structures. It appears from our
analysis that weak corrugation of the guiding channel can add
unexpected richness to this picture. The addition of transient
effects, binary interaction between droplets and surfactants seems
likely to add even more to this richness. Exploring these effects
appears promising both for microfluidic applications and for basic
understanding of what capillary flow can do.\\

\begin{acknowledgments}
\indent The authors are grateful to Prof. Rustem Ismagilov and Dr.
Ying Liu, whose microfluidic experiments inspired this work. This
work was supported in part by the National Science Foundation's
MRSEC Program under Award Number DMR 0820054. And by a grant from
the US-Israel Binational Science Foundation.\\
\end{acknowledgments}

\appendix

\section{}

\indent In this appendix, we derive the dependence of the rescaled
curvature $P$ on the finite plug length $\xi_p$. All quantities are
in the dimensionless units defined by the rescalings in Section 2.
To set the stage for our treatment of finite plugs, we analyze the
behavior of $\eta$ for the semi-infinite plug treated by Bretherton.
We proceed from a semi-infinite plug with the front cap. Far inside
the thin-film region, $|\eta -1| \ll 1$ and we may use the general
solution in Eqn.(\ref{eqn:1:generalsolution}). In the front region
of the thin film, the contributions from $\psi_2$ and $\psi_3$ are
arbitrarily small relative to that from $\psi_1$. So the plug shape
takes the form:
\begin{equation}
\eta = 1 + C_1 \psi_1(\xi) \, . \label{eqn:1:psi1only}
\end{equation}
\noindent Consistency with this form implies: $\eta_{\xi\xi} =
\eta_{\xi} = \eta - 1$, where $\eta_{\xi\xi}$ and $\eta_{\xi}$ stand
for the second and the first derivative of $\eta$ with respect to
$\xi$ respectively. We may thus find the profile $\eta(\xi)$ by
integrating forward from an arbitrary origin to $\xi \rightarrow
+\infty$, starting from an initial value of $\eta$ very close to
$1$. The asymptotic curvature $\eta_{\xi\xi}(+\infty)$ is
necessarily finite and independent of axial translation in the
small-amplitude limit. Specifically, $\eta_{\xi\xi} (+\infty)
\rightarrow P_\infty = 0.643$. It is convenient to define $\xi_F$ as
the $\xi$ for which the $\psi_1$ contribution in
Eqn.(\ref{eqn:1:psi1only}) extrapolates to 1, \textit{i.e.} $C_1
\psi_1(\xi_F)= 1$. Thus the thin film region of
Eqn.(\ref{eqn:1:psi1only}) is where $\xi_F - \xi \gg 1$, and
$\eta(\xi) = 1 + e^{\xi - \xi_F}$. Then from the $\eta(\xi)$ profile
thus determined numerically, one finds that Teletzke's point $B$ has
position: $\xi_B = \xi_F - 0.17$. Evidently,
$\eta_{\xi\xi}(+\infty)$ is independent of $\xi_F$. \\

\indent A similar analysis can be applied to a semi-infinite plug
with the rear cap. Here, the $\psi_1$ contribution is negligible in
the rear region of the thin film. So the plug shape takes the form:
\begin{equation}
\eta = 1 + C_2 \psi_2(\xi) + C_3 \psi_3(\xi) \, .
\end{equation}
\noindent As with the front region, we may find the rear $\eta$
profile by integrating backward from the thin-film region where
$\eta \rightarrow 1$. Far inside the thin-film region the $\psi_2$
and $\psi_3$ contributions are necessarily small, but their ratio
varies sinusoidally with position, with a wavelength comparable to
the remobilization length $L_\infty$. As above, we define a $\xi_R$
such that $C_2 \psi_2(\xi_R) = 1$. Then in the thin film region
$\eta$ takes the form:
\begin{equation}
\eta = 1 +
e^{-(\xi-\xi_R)/2}\cdot\left[\cos{(\sqrt{3}(\xi-\xi_R)/2)}+C_3
\,\sin {(\sqrt{3}(\xi-\xi_R)/2)}\right] \, . \label{eqn:1:C2equal1}
\end{equation}
\noindent To determine $\eta(\xi)$ in the nonlinear regime, we
integrate towards negative $\xi$ starting from initial conditions
compatible with this form. An initial value $\xi_0$ is chosen such
that $\eta(\xi_0)$ is very close to $1$. Compatibility with
Eqn.(\ref{eqn:1:C2equal1}) dictates the values of $\eta_\xi(\xi_0)$
and $\eta_{\xi\xi}(\xi_0)$. Then we integrate towards $\xi
\rightarrow -\infty$ to obtain the profile. As with the front
profile, the asymptotic rear curvature $\eta_{\xi\xi}(-\infty)$ is
necessarily finite and independent of translation in $\xi$. However,
this $\eta_{\xi\xi}(-\infty)$ depends on the choice of $C_3$ values.
The correct profile is that which matches the front curvature
$\eta_{\xi\xi}(-\infty) = \eta_{\xi\xi}(+\infty)=0.643$. This
requirement fixes the value of $C_3$ in Eqn.(\ref{eqn:1:C2equal1}).
We find numerically that $C_3 = -0.85$. As with the front profile,
we may determine the rear Teletzke point $\xi_A$ by numerically
solving the quadratic coefficients of Eqn.(\ref{eqn:1:parabolas})
and get: $\xi_A =
\xi_R+3.37$. \\

\indent We now focus on long plugs of finite length: $\xi_p \gg 1$.
To the lowest order approximation, the above determination of
Teletzke points allows us to relate the dimensionless plug length
$\xi_p$ to $\xi_F - \xi_R$ as follows:
\begin{eqnarray}
\xi_F - \xi_R &=& \xi_p + 0.17 + 3.37 \nonumber \\
&=& \xi_p + 3.54 \, . \label{eqn:1:xiF-xiR}
\end{eqnarray}
\noindent Obviously $\xi_F - \xi_R \gg 1$ in the regime of interest.
For finite plugs, the thin film region contains nonzero
contributions from all three modes $\psi_1$, $\psi_2$, and $\psi_3$.
Using our conventions for the $\psi$'s above, $\eta$ takes the
following form in the front region of the thin film:
\begin{eqnarray}
\eta &=& 1 + e^{\xi-\xi_F} +
e^{-(\xi-\xi_R)/2}\cdot\left[\cos{(\sqrt{3}(\xi-\xi_R)/2)}-0.85
\,\sin {(\sqrt{3}(\xi-\xi_R)/2)}\right] \nonumber \\
&=& 1 + \psi_1(\xi-\xi_F) + \psi_2(\xi - \xi_R) - 0.85\,
\psi_3(\xi-\xi_R) \nonumber \\
&=& 1 + \psi_1(\xi-\xi_F) + \psi_2(\xi - \xi_F + \xi_F - \xi_R)
-0.85\, \psi_3(\xi - \xi_F + \xi_F - \xi_R) \, ,
\label{eqn:1:etafiniteplug}
\end{eqnarray}
\noindent where higher order perturbations due to finite plug length
are neglected. The admixture of $\psi_2$ and $\psi_3$ contributions
alter the behavior at large $\xi$ and thus perturbs the asymptotic
curvature $P$. In order to find this perturbation, we must first
express the $\psi$'s using a common origin $\xi_F$. By using the
addition properties of sine and cosine functions, we have:
\begin{eqnarray}
\psi_2(\xi - \xi_F + \xi_F - \xi_R) & = & \psi_2(\xi -
\xi_F)\psi_2(\xi_F - \xi_R) - \psi_3(\xi - \xi_F)\psi_3(\xi_F -
\xi_R)
\nonumber \\
\psi_3(\xi - \xi_F + \xi_F - \xi_R) & = & \psi_2(\xi -
\xi_F)\psi_3(\xi_F - \xi_R) + \psi_3(\xi - \xi_F)\psi_2(\xi_F -
\xi_R) \, .
\end{eqnarray}
\noindent Plugging the above expressions into
Eqn.(\ref{eqn:1:etafiniteplug}) and using Eqn.(\ref{eqn:1:xiF-xiR})
to substitute $\xi_F - \xi_R$, we get:
\begin{equation}
\eta = 1 + \psi_1(\xi-\xi_F) + \delta_2(\xi_p) \cdot \psi_2(\xi -
\xi_F) + \delta_3(\xi_p) \cdot \psi_3(\xi - \xi_F) \, ,
\label{eqn:1:etawithdelta23}
\end{equation}

\noindent where $\delta_2$ and $\delta_3$ are functions of the plug
length $\xi_p$ defined as follows:
\begin{equation}
\left\{\begin{array}{c}
\delta_2(\xi_p) \\
\delta_3(\xi_p)
\end{array}\right\}
\equiv \left(\begin{array}{cc}
1 & -0.85 \\
-0.85 & -1
\end{array}\right) \cdot
\left\{\begin{array}{c}
\psi_2(\xi_p + 3.54) \\
\psi_3(\xi_p + 3.54)
\end{array}\right\} \, _. \label{eqn:1:delta23exp}
\end{equation}
\noindent In the regime of interest $\xi_p \gg 1$, $\delta_2$ and
$\delta_3$ are very small: $|\delta_{2,3}(\xi_p)|\ll 1$. Infinite
plugs correspond to the limit case where $\delta_{2,3}(+\infty)
\rightarrow 0$. Similar to the above analysis, we integrate towards
$\xi \rightarrow +\infty$ from some point $\xi_0$ far inside the
thin film region with initial conditions compatible with
Eqn.(\ref{eqn:1:etawithdelta23}). The asymptotic front curvature
$\eta_{\xi\xi}(+\infty)$ is necessarily finite and independent of
small-amplitude translation in $\xi$. However, this
$\eta_{\xi\xi}(+\infty)$ depends on the plug length $\xi_p$.
Moreover, it is evident from the functional form of $\eta$ in
Eqn.(\ref{eqn:1:etawithdelta23}) that $\eta_{\xi\xi}(+\infty)$ has
dependence on $\xi_p$ only through functions $\delta_2(\xi_p)$ and
$\delta_3(\xi_p)$. By defining $P\equiv\eta_{\xi\xi}(+\infty)$, we
have:
\begin{eqnarray}
P(\xi_p) &=& P(\delta_2(\xi_p),\delta_3(\xi_p)) \nonumber \\
&\simeq& P(0,0)+\left.\delta_2(\xi_p)\cdot\frac{\partial P}{\partial
\delta_2}\right|_{(0,0)} +\left.\delta_3(\xi_p)\cdot\frac{\partial
P}{\partial \delta_3}\right|_{(0,0)} \nonumber \\
&=& P_{\infty} + \left.\delta_2(\xi_p)\cdot\frac{\partial
P}{\partial \delta_2}\right|_{(0,0)} +
\left.\delta_3(\xi_p)\cdot\frac{\partial P}{\partial
\delta_3}\right|_{(0,0)} \, , \label{eqn:1:Pexpansion}
\end{eqnarray}
\noindent where we have used the fact that $\delta_2$ and $\delta_3$
take vanishingly small values in the regime of interest and kept
only the lowest order approximation. Partial derivatives $(\partial
P/\partial \delta_2)|_{0,0}$ and $(\partial P/\partial
\delta_3)|_{0,0}$ are sensitivity factors that we can determine
numerically as follows:
\begin{eqnarray}
\left.\frac{\partial P}{\partial \delta_2}\right|_{(0,0)} &=& -1.38
\, ,
\nonumber \\
\left.\frac{\partial P}{\partial \delta_3}\right|_{(0,0)} &=& 0.48
\, .
\end{eqnarray}
\noindent Plugging these sensitivity factors into
Eqn.(\ref{eqn:1:Pexpansion}), we get:
\begin{equation}
P(\xi_p) = P_\infty -1.38 \,\delta_2(\xi_p) + 0.48 \,\delta_3(\xi_p)
\, . \label{eqn:1:Pnumeric}
\end{equation}
\noindent This is the dependence of the curvature $P$ on the plug
length $\xi_p$ stated in the text.\\

\label{lastpage}


\begin{thebibliography}{13}
\bibitem[(Bretherton 1961)]{Bretherton_1961}\textsc{Bretherton, F. P.} 1961, \emph{J. Fluid Mech.}, \textbf{10}, 166.

\bibitem[(Ismagilov \& Ying)]{Ying_2008} \textsc{Ismagilov, R. \& Liu, Y.}, \emph{Private Communication}.

\bibitem[(Gunther \& Jensen 2006)]{Gunther_2006} \textsc{Gunther, A. \& Jensen, K. F.} 2006, \emph{Lab Chip}, \textbf{6}, 1487.

\bibitem[Stillwagon \& Larson(1988)]{Stillwagon_1988} \textsc{Stillwagon, L. E. \& Larson, R. G.} 1988, \emph{J. Appl. Phys.} \textbf{63}, 5251.

\bibitem[Schwartz \& Weidner(1995)]{Schwartz_1995} \textsc{Schwartz, L. W. \& Weidner, D. E.} 1995, \emph{J. Engng. Math.}, \textbf{29}, 91.

\bibitem[Kalliadasis, Bielarz \& Homsy(2000)]{Kalliadasis_2000} \textsc{Kalliadasis, S., Bielarz, C., \& Homsy, G. M.} 2000, \emph{Phys. Fluids}, \textbf{12}, 1889.

\bibitem[Mazouchi \& Homsy(2001)]{Mazouchi_2001} \textsc{Mazouchi, A. \& Homsy, G. M.} 2001, \emph{Phys. Fluids}, \textbf{13}, 2751.

\bibitem[Howell(2003)]{Howell_2003} \textsc{Howell, P. D.} 2003, \emph{J. Engng. Math.}, \textbf{45}, 283.

\bibitem[(Ajaev \& Homsy 2006)]{Ajaev_2006} \textsc{Ajaev, V. S. \& Homsy, G. M.} 2006, \emph{Annu. Rev. Fluid. Mech.}, \textbf{38}, 277.

\bibitem[(Chen 1986)]{Chen_1986} \textsc{Chen, J. D.} 1986, \emph{J. Coll. Interf. Sci.}, \textbf{109}(2), 341.

\bibitem[Einzel, Panzer \& Liu(1990)]{Einzel_1990} \textsc{Einzel, D., Panzer, P. \& Liu, M.} 1990, \emph{Phys. Rev. Lett.}, \textbf{64}, 2269.

\bibitem[Miksis \& Davis(1994)]{Miksis_1994} \textsc{Miksis, M. J. \& Davis, S. H.} 1994, \emph{J. Fluid Mech.}, \textbf{273}, 125.

\bibitem[DeGennes(2002)]{DeGennes_2002} \textsc{de Gennes, P. G.} 2002, \emph{Langmuir}, \textbf{18}, 3413.

\bibitem[(Krechetnikov \& Homsy 2005)]{Krechetnikov_2005} \textsc{Krechetnikov, R. \& Homsy, G. M.} 2005, \emph{Phys. Fluids}, \textbf{17}, 102108.

\bibitem[Wong$^1$(1995))]{Wong_1995} \textsc{Wong, H., Radke, C. J. \& Morris, S.} 1995, \emph{J. Fluid Mech.}, \textbf{292}, 71.

\bibitem[Wong$^2$(1995))]{Wong_1995_2} \textsc{Wong, H., Radke, C. J. \& Morris, S.} 1995, \emph{J. Fluid Mech.}, \textbf{292}, 95.

\bibitem[(Hazel \& Heil 2002)]{Heil_2002} \textsc{Hazel, A. L. \& Heil, M.} 2002, \emph{J. Fluid Mech.}, \textbf{470}, 91.

\bibitem[Wenzel(1936)]{Wenzel_1936} \textsc{Wenzel, R. N.} 1936, \emph{Ind. Eng. Chem.}, \textbf{28}, 988.

\bibitem[Herminghaus(2000)]{Herminghaus_2000} \textsc{Herminghaus, S.} 2000, \emph{Europhys. Lett.}, \textbf{52}, 165.

\bibitem[Bico, Tordeux \& Quere(2001)]{Bico_2001} \textsc{Bico, J., Tordeux, C. \& Quere, D.} 2001, \emph{Europhys. Lett.}, \textbf{55}, 214.

\bibitem[(Teletzke 1983)]{Teletzke_1983} \textsc{Teletzke, G. F.} 1983, \emph{Ph.D. Thesis}, Univeristy of Minnesota.

\bibitem[(Hodges, Jensen \& Rallison 2004)]{Jensen_2004} \textsc{Hodges, S. R., Jensen, O. E. \& Rallison, J. M.} 2004, \emph{J. Fluid Mech.}, \textbf{501}, 279.

\bibitem[(DeGennes, Brochard \& Quere 2003)]{deGennes_2003} \textsc{de Gennes, P. G., Brochard, F. \& Quere, D.} \emph{Capillarity and Wetting Phenomena}
(Springer-Verlag, New York, 2003).

\bibitem[(Brenner 1963)]{Brenner_1963} \textsc{Brenner, H.} 1963, \emph{Chem. Eng. Sci}, \textbf{19}, 519.

\end{thebibliography}
\end{document}